\documentclass{svjour3}                    
\smartqed  
\usepackage{graphicx}
\usepackage{amssymb}
\usepackage{amsfonts}
\usepackage{enumerate} 
\usepackage{array}
\usepackage{amsmath}
\usepackage{comment}

\usepackage{natbib}

\newcommand{\eps}{\varepsilon}
\newcommand{\e}{{\rm e}}

\journalname{Celestial Mechanics and Dynamical Astronomy}

\begin{document}

\title{High precision Symplectic Integrators for the Solar System
}


\author
{
	Ariadna Farr\'es 	\and
	Jacques Laskar   	\and 
	Sergio Blanes    	\and 
	Fernando Casas 	\and 
	Joseba Makazaga   \and
	Ander Murua      
}


\institute{
	Ariadna Farr\'es \and Jacques Laskar \at
    Astronomie et Syst\`emes Dynamiques, IMCCE-CNRS UMR8028, Observatoire de Paris, UPMC,, 77 Av. Denfert-Rochereau, 75014-Paris, France 
	\email{afarres@imcce.fr,\ laskar@imcce.fr}
    \and
   Sergio Blanes \at
	Instituto de Matem\'atica Multidisciplinar, Universitat Polit\'ecnica de Val\`encia, 
	46022-Valencia, Spain 
	\email{serblaza@imm.upv.es}
	\and
	Fernando Casas \at 
	Institut de Matem\`atiques i Aplicacions de Castell\'o and Departament de
	Matem\`atiques, Universitat Jaume I, E-12071 Castell\'on, Spain 
	\email{Fernando.Casas@uji.es} 
	\and 
	Joseba Makazaga \and Ander Murua \at
	Konputazio Zientziak eta A.A. saila, Informatika Fakultatea, EHU/UPV,
	Donostia/San Sebasti\'an, Spain \email{Joseba.Makazaga@ehu.es,\ Ander.Murua@ehu.es}
}

\date{Received: date / Accepted: date}

\maketitle

\begin{abstract}

Using a Newtonian model of the Solar System with all 8 planets, we perform 
extensive tests on various symplectic integrators of high orders, searching 
for the best  splitting scheme for long term studies in the Solar System.
These comparisons are made in Jacobi and Heliocentric coordinates and 
the implementation of the algorithms is fully detailed for practical use. 
We conclude that high order integrators should be privileged, with 
a preference for the new $(10,6,4)$ method of \citep{Bla12}.

\end{abstract}

\section{Introduction}
\label{intro}

Due to their simplicity and  stability properties, symplectic integrators have
been widely used for long-term integrations of the Solar System, starting with
the work of~\cite{WiHo91}. In many studies on the formation and evolution of the
Solar System, where large numbers of particles are considered, the speed of the
integrator is a major constraint and low order schemes have been  often
privileged as in the original scheme of  \citep{WiHo91} or \citep{KiYoNa91} (for
a review see \citep{Morb2002a}).

On the opposite, in the present work we are focusing on high precision
symplectic integrators that are designed for the computation of long term
ephemerides of the Solar System, when one searches to reduce the numerical
error of the algorithm  to the level of the  roundoff error of the  machine.
 These integrators will also be useful for the detailed dynamical studies of the
extra solar  planetary system with strong planetary interactions.

The first long term direct numerical integration of a realistic model of the
Solar System, including all planets and the effects of general relativity and
the Moon was made twenty years ago over 3 Myr \citep{QuinTrem1991a} using a high
order symmetric multistep method. This solution could be compared with success 
with the previous  averaged solutions of \citep{Lask1989a,Lask1990a} and
confirmed the existence of secular resonances in the Solar System
\citep{LaskQuin1992a}. Soon after, using a symplectic integrator with mixed
variables \citep{WiHo91}, \citet{SussWisd1992a} could extend these computation
to 100 Myr, confirming the chaotic behaviour of the Solar System discovered with
the secular equations  by  \citet{Lask1989a,Lask1990a}.

As the Solar System is chaotic, the error in numerical integrations is
multiplied by 10 every 10 Myr \citep{Lask1989a}. Due to the limited accuracy of
the models and initial conditions, it is thus hopeless to obtain a  precise
solution for the evolution of the Solar System over  more than 100 Myr. The
situation is even worse when the full Solar System is considered, as close
encounters among the minor planets induce strong chaotic effects that will limit
all possibilities of computing a precise solution for the planets to about 60
Myr \citep{LaskFien2011a,LaskGast2011a}.

Despite this limitation, there is a strong  need for precise ephemerides of the
planets from the paleoclimate community. Indeed, the variations of the Earth
orbital elements induce some changes in the Earth climate that are reflected in
the sedimentary records over million of years. This mechanism, known as
Milankovitch theory \citep{Mila1941a} allows now to use the astronomical
solution for the calibration of the geological time scales through the
correlation of the variation of orbital and rotational elements of the Earth to
geological records. This method has been successfully used for the Neogene
period  \citep{L.J.Wils2004a} over 23 Myr, and a large effort is pursued at
present to extend this study over the full  Cenozoic era, up to about 65 Myr.
This quest led to search for high order symplectic schemes that are adapted to
these long time computations, where high accuracy is requested
\cite{LaRo01,LaskRobu2004a,LaskFien2011a}, but it should be noted that in the
latest work, the integration of the Solar System model over 250
Myr\footnote{Although it has been demonstrated that  a precise solution of the
motion of the Earth cannot be computed over more than 60 Myr
\citep{LaskGast2011a}, the solutions are systematically computed over 250 Myr as
some features of the solutions can be trusted over longer times
\cite{LaskRobu2004a,LaskFien2011a}.}, including five main asteroids took more
than 18 months of CPU time. Some improvements of the algorithms were thus
needed, and the present paper is the outcome of the studies that we have
understaken in order to search for the best integrators for the next generations
of numerical solutions. At the same time, we have compared various sets of
coordinates (Heliocentrics  and Jacobi), as the performances of these
integrators depend on the choice of splitting of the Hamiltonian, and thus of
the set of coordinates that correspond to these various splittings. As the 
integrators  that are presented here are of high order, they can also be used
for refined analysis of the newly discovered extra solar  planetary systems,
especially when the planetary interactions in the system are strong.

For the planetary case, when using an appropriate set of
coordinates, the equations of motion are written as an integrable part $H_A$,
that corresponds to the Keplerian motion of each planet, and a small
perturbation $H_B$, given by the interaction of the planets between each other.
Hence, the system falls into the category of Hamiltonian system of the kind $H
= H_A + \eps H_B$. 

Several  splitting integrating schemes that take advantage 
of this fact to derive efficient integrators exist in the literature. 
\cite{McLa9502} was the first to present 
such schemes and was  followed independently by \cite{Cham00} and \cite{LaRo01}.
Most recently works by \cite{Bla12} derived higher order schemes that present 
very interesting behaviour. 

In this paper we describe these different splitting symplectic schemes and
compare them for the case of the Solar System dynamics. We want to see which
are the most efficient and accurate schemes. We will consider the gravitational
N-body model and test the different integrating schemes against different
planetary configurations, to be more specific: the 4 inner planets, the 4 outer
planets and the 8 planets in the Solar System (Section~\ref{TEST_MODEL}). 

The Hamiltonian of the gravitational N-body problem $H = T(p) + U(q)$ can be
rewrite as $H = H_A + \eps H_B$, using to different set of canonical
coordinates: Jacobi and Heliocentric coordinates (Section~\ref{NBP_coord}).
The main difference between both set of equations is that in Jacobi coordinates
the small perturbation $H_B$ depends only in positions while in Heliocentric
coordinates these one depends on both position and velocity. This is why in the
literature Jacobi coordinates have been more widely used. In
Section~\ref{SymIntTST} we describe different symplectic schemes for Jacobi
coordinates, and in Section~\ref{SymIntHelio} other symplectic schemes that are
suitable for Heliocentric coordinates. In both Sections we describe and
compare the different splitting schemes. Finally in Section~\ref{SymIntHvsJ} we
compare the results for the two different set of coordinates.

\section{Splitting Symplectic Integrators (General Overview)} \label{SplSym_Intro}

Let $H(q,p)$ be a Hamiltonian system where $(q,p)$ are a set of canonical
coordinates (i.e. $q$ are the positions and $p$ the momenta). It is well known
that in many mechanical problems the Hamiltonian is of the form $$H(q,p) =
T(p) + U(q),$$ which is separable with respect to the local canonical coordinates. 
Using the Lie formalism 
we can write the equations  of motion as: 
\begin{equation}
\frac{d z}{dt} = \{H, z\} = L_H z, 
\label{EqHam01}
\end{equation}
where by definition $L_\chi f := \{\chi,f\}$ is the differential operator 
$L_\chi$, $z = (q,p)$ and $\{\cdot,\cdot\}$ denotes the Poisson 
bracket\footnote{$\{F,G\} = \sum_{i=1}^n
	\frac{\partial F}{\partial p_i}\frac{\partial G}{\partial q_i} 
	- \frac{\partial F}{\partial q_i}\frac{\partial G}{\partial p_i}$}.

The formal solution of Eq.~\ref{EqHam01} at time $t = \tau_0 + \tau$ that starts at time
$t = \tau_0$ is given by 
\begin{equation}
z(\tau_0 + \tau) = \exp(\tau L_H)z(\tau_0) = \exp(\tau (L_T+L_U))z(\tau_0).
\label{EqSolHam01}
\end{equation} 

In general the operators $L_T$ and $L_U$ do not commute, $\exp(\tau (L_T+L_U))
\neq \exp(\tau L_T) \exp(\tau L_U)$, but we can find coefficients $a_i, b_i$
such that for a given $r$, 
\begin{equation}
	\exp(\tau (L_T+L_U)) = \displaystyle\prod_{i=1}^{s} 
	 \exp(a_i\tau L_T) \exp(b_i\tau L_U) + O(\tau^{r+1}). 
\label{SplitIdea01}
\end{equation}
	
Using the Baker--Campbell--Hausdorff (BCH) identity we can find relations
that the coefficients $a_i, b_i$ must satisfy to have a high order
scheme~\citep{Kose93,Kose96}. These are the so-called {\bf order conditions}.
For a given set of coefficients $a_i, b_i$ 
satisfying Eq.~\ref{SplitIdea01}, the composition 
\begin{equation}	
	z(\tau) = \mathcal{S}(\tau) z(\tau_0) = 
	\displaystyle\prod_{i=1}^{s} \exp(a_i\tau L_T) \exp(b_i\tau L_U)z(\tau_0),
	\label{SplitIntBase}
\end{equation}
is a symplectic map of order $r$. 

The map $\mathcal{S}(\tau)$ is symplectic because it is the product of
elementary symplectic maps, $\exp(\tau L_T)$ and $\exp(\tau L_U)$, and has
order $r$ because it approximates the exact solution up to order $\tau^r$.
We will refer to these kind of symplectic schemes as { splitting symplectic
integrators}. 

Some of the main advantages of these kind of integrating schemes are: a) they
are very easy to implement; b) they preserve the symplectic character of the
Hamiltonian system; and c) in general there is no systematic drift on the
conservation of the energy during the numerical integration.

These kind of symplectic schemes have been widely studied throughout the years
by several authors (see \cite{hairer06gni,mclachlan02sm} and references
therein). As a matter of fact, splitting methods have been designed (often
independently) and extensively used in fields as diverse as molecular dynamics,
simulations of storage rings in particle accelerators, quantum chemistry and, of
course, celestial mechanics.

There are several procedures to get the order conditions for the coefficients of
the splitting scheme in Eq.~\ref{SplitIntBase}. These are, generally speaking,
large systems of polynomial equations in the coefficients that are obtained from
Eq.~\ref{SplitIdea01}. Two of the most popular are the recursive  application of the 
BCH formula to the composition in Eq.~\ref{SplitIntBase}, and a generalisation of the 
theory of rooted trees used in the analysis of Runge–Kutta methods due to 
\cite{Murua15041999} (see also \cite{hairer06gni}).
The later procedure, while being more systematic than the former, is however 
not appropriate for the case splitting methods applied to Hamiltonians of the 
form $A+\epsilon B$. In \cite{Bla12} a novel systematic way is proposed 
based on the so-called Lyndon multi-indices that is well adapted to that case.

Splitting methods of order greater than two involve necessarily some negative
coefficients $a_i$ and $b_j$ \citep{goldman96noo,sheng89slp,Suzu91}. Although
this feature does not imply in principle any special impediment for the class of
systems considered in this paper, it is clear that the presence of negative
coefficients may affect the numerical error and the maximal step size of the
scheme. For this reason, when dealing with high order methods, minimising the
size of the negative coefficients and the sum of the absolute value of all the
coefficients will be a critical factor in the choice of one particular set of
coefficients. 

%

In this paper we do not intend to give the details on the derivation of the order 
conditions or how to find these coefficients. All these issues are analysed in detail 
in \cite{Bla12}. Our aim here is to compare the performance of different splitting 
symplectic schemes for the specific case of the integration of 
the Solar System. 

If we focus on the motion of the Solar System, or other planetary systems, 
we have a main massive body in the centre (the Sun) and the other bodies evolve
around the centre mass following almost Keplerian orbits. We can take advantage
of this to build more efficient schemes. Using an appropriate change of
coordinates we can rewrite the Hamiltonian as, $H = H_{K} + H_{I}$ (where 
$|H_I| \ll |H_K|$), a sum of the Keplerian motion of each planet around the 
central star and a small perturbation due to the interaction between the planets,  
where $H_{K}$ and $H_{I}$ are integrable.

\cite{WiHo91,KiYoNa91} where the first to split the Hamiltonian of the N-body
problem in this way for numerical simulations of the Solar System, by means of what 
is called a mixed variable integrator, using elliptical coordinates to integrate 
the Keplerian motion. Splitting the Hamiltonian as $H_{K} + H_{I}$ rather than the 
classical $T(p) + U(q)$ already improves the performance 
of the leapfrog scheme. As $|H_{I}|$ is small with respect to $|H_{K}|$, the 
system falls into the class of Hamiltonian such
that $H = H_A + \eps H_B$ for $\eps$ small. In this particular case, the
truncation order of the leapfrog scheme is no longer $C\tau^3$ as for $T(p) +
U(q)$, but rather $C' \eps\tau^3$ \citep{McLa9502,LaRo01}.  

In Sections~\ref{SymIntTST} and~\ref{SymIntHelio} we will describe different 
families of symplectic splitting methods for Hamiltonian systems of the kind 
$H_A + \eps H_B$ and we will compare their performance for the particular 
case of the Solar System dynamics. 

\section{The N-Body Problem}\label{NBP_coord}

Throughout this article we consider the non-relativistic gravitational N-body
problem as a test model for the different integrating schemes. We are aware
that to have a realistic model for the Solar System dynamics one must include 
effects like general relativity or tidal dissipation. Nevertheless, and 
for the sake of simplicity, in this paper these effects are ignored as their 
presence should not compromise the performance of the schemes presented here. 

In a general framework, we consider the motion of $n+1$ particles: the Sun
and $n$ planets, that are only affected by their mutual gravitational
interaction. Let $\bf u_0, u_1, \dots, u_n$ and $\bf \dot u_0, \dot u_1, \dots,
\dot u_n$ be the position and velocities, in a barycentric reference frame, of
the $n+1$ bodies and let $m_0, m_1, \dots, m_n$ be their respective masses. For
simplicity, we consider $m_0$ to be the mass of the Sun and $m_i$ for $i=1,
\dots, n$ the mass of the other planets. 

Taking the conjugated momenta ${\bf \tilde u_i} = m_i {\bf \dot u_i}$, the
equations of motion are Hamiltonian, with:  
\begin{equation} 
H = \frac{1}{2}\sum_{i=0}^{n}\frac{|| {\bf \tilde u_i} ||^2}{m_i}
    - G\sum_{0\leq i < j \leq n} \frac{m_im_j}{|| {\bf u_i - u_j} ||}.
\label{HBari}
\end{equation}
In this set of coordinates the Hamiltonian naturally splits into, $H = T + U$,
where $T$ depends only on the momenta (${\bf \tilde u_i}$) and $U$ depends only
on the positions (${\bf u_i}$). 

In general, when we deal with complex dynamical systems, it is important to
take into account the relevant aspects of the system and use them to build
efficient numerical tools to describe their dynamics. In
the case of the Solar System we have a massive body in the centre and the
planets evolve following Keplerian orbits around it that vary through time
due to their mutual interaction. 

Using an appropriate change of variables the Hamiltonian can be written as 
$H_{K} + H_{I}$, where $|H_{I}|$ is small with respect to $|H_{K}|$, and 
both parts are integrable when we considered them on their own. There exist two 
canonical set of coordinates that allow us to split the Hamiltonian in this 
way: Jacobi and Heliocentric coordinates. 

\subsection{Jacobi Coordinates}\label{JacobCoord}

The Jacobi set of coordinates have been widely used in Celestial Mechanics for 
developing analytical theories for the planetary motion. They where first 
used for the numerical integration of the Solar System by \cite{WiHo91}. 

Here the position of each planet, ${\bf v_i}$ for $i=1,\dots,n$, is considered 
relative to the barycentre ${\bf G_{i-1}}$ of the previous $i$ bodies, 
${\bf u_0, \dots, u_{i-1}}$, and ${\bf v_0}$ is taken as the centre of mass 
of the system: 
\begin{equation}
\left.\begin{array}{lcl}
\bf v_0 &=& (m_0{\bf u_0} + \dots + m_n{\bf u_n})/\eta_n  \\
\bf v_i &=& {\bf u_i} - (\sum_{j=0}^{i-1} m_j{\bf u_j})/\eta_{i-1} 
\end{array}\right\}, 
\label{Bar2Jac00}
\end{equation} 
where $\eta_i = \sum_{j = 0}^{i} m_j$. To have a canonical change of variables 
the momenta ${\bf \tilde v_i}$ for $i=0,\dots,n$, must be: 
\begin{equation}
\left.\begin{array}{lcl}
\bf \tilde v_0 &=& \bf \tilde u_0 + \dots + \tilde u_n  \\
\bf \tilde v_i &=& (\eta_{i-1}{\bf \tilde u_i} - m_i\sum_{j=0}^{i-1} {\bf \tilde u_j})/\eta_{i}
\end{array}\right\}.
\label{Bar2Jac11}
\end{equation}
In this set of coordinates the Hamiltonian in Eq.~\ref{HBari} takes the form
\citep{CursLaskar}: 
\begin{equation}
\begin{array}{rcl}
H_{Jb} &=& 
  \displaystyle
	\sum_{i=1}^{n} \left( 
  	\frac{1}{2} \frac{\eta_i}{\eta_{i-1}}\frac{||{\bf \tilde v_i}||^2}{m_i} 
  	- G\frac{m_i \eta_{i-1}}{||{\bf v_i}||} 
  \right) \\ 
&& \\
 &+& \displaystyle G \left[ \sum_{i=2}^{n}m_i\left(\frac{\eta_{i-1}}{||{\bf v_i}||} 
  - \frac{m_0}{||{\bf r_i}||}\right) 
  - \sum_{0 < i < j\leq n} \frac{m_i m_j}{\Delta_{ij}} \right], 
 \end{array}
\label{HJacob}
\end{equation}
where $\Delta_{i,j} = || {\bf u_i - u_j} ||$ (the distance between the two
bodies) can be expressed as a function of ${\bf v_i}$ and ${\bf v_j}$, and
${\bf r_i = u_i - u_0}$. If we fix the centre of mass at the origin
then $\bf v_0 = 0$ and $\bf \tilde v_0 = 0$, and we reduce in $6$ the number of
equations of motion. 

\subsection{Heliocentric Coordinates}\label{HelioCoord} 

Here we consider the relative position of each planet with respect to the Sun: 
\begin{equation}
\left.\begin{array}{lcl}
\bf r_0 &=& \bf u_0  \\
\bf r_i &=& \bf u_i - u_0
\end{array}\right\}, 
\label{Bar2Hel00}
\end{equation}
and to have a canonical change of variables the momenta are: 
\begin{equation}
\left.\begin{array}{lcl}
\bf \tilde r_0 &=& \bf \tilde u_0 + \dots + \tilde u_n  \\
\bf \tilde r_i &=& \bf \tilde u_i
\end{array}\right\}.
\label{Bar2Hel}
\end{equation}
In this set of coordinates the Hamiltonian in Eq.~$\ref{HBari}$ takes the 
form \citep{CursLaskar}: 
\begin{equation}
H_{He} = 
  \sum_{i=1}^{n} \left( 
  \frac{1}{2} ||{\bf \tilde r_i}||^2 \left[\frac{m_0 + m_i}{m_0 m_i}\right] 
  - G\frac{m_0 m_i}{||{\bf r_i}||} 
  \right)
  + \sum_{0 < i < j\leq n}\left(
  \frac{\bf \tilde r_i \cdot \tilde r_j}{m_0} - G\frac{m_i m_j}{\Delta_{ij}}
  \right), 
\label{HHelio}
\end{equation}
where $\Delta_{i,j} = || {\bf r_i - r_j} ||$ for $i,j>0$. 
If we consider the centre of mass of the system to be fixed at the
origin we have that ${\bf \tilde r_0} = 0$, and we can easily recompute 
$\bf r_0$ at all time. Hence, we have also reduced in 6 the number of
equations of motion. 

One of the main differences between these two sets of coordinates is the size of 
the perturbation which in the case of Jacobi coordinates is smaller than 
for Heliocentric coordinates (see Table~\ref{TPertSize} in Section~\ref{TEST_MODEL}). 

Moreover, in the case of Jacobi coordinates the perturbation part ($H_I$) depends only on
positions so it is integrable when we consider it alone. But the expressions
are more cumbersome than for Heliocentric coordinates (see
Appendix~\ref{IntScheme}). While in the case of Heliocentric coordinates the
perturbation part depends on both position and velocities, hence it is not
integrable on its own. In Section~\ref{SymIntHelio} we will show how to adapt
the splitting schemes to this particular case.

\section{Test to Perform} \label{TEST_MODEL}

Let $S(\tau) = \prod_{i=1}^s \exp(a_i\tau A)\exp(b_i\tau B)$ be a splitting
symplectic scheme.  We say $S(\tau)$ has {\bf $s$ stages} if it requires $s$
evaluations of $\exp(\tau A)\exp(\tau B)$ per step-size. The smaller the
step-size $\tau$ used, the smaller is the error of the numerical solution
provided by the scheme, and the larger is the computational cost, as more
evaluations of  $\exp(\tau A)\exp(\tau B)$ are required to integrate over the
same time period.

Usually, the higher the order of the scheme the more number $s$ of stages it
requires, increasing the computational cost to advance a given step-size $\tau$.
So a method with 4 stages will be more efficient than one with 2 stages if it
can integrate a given accuracy with a step-size which is at least two times
larger than the one required for the 2 stages scheme. In this sense, we define
the { inverse cost} of $S(\tau)$ as $\tau/s$, where $s$ is the number of
stages and $\tau$ is the step-size used. Thus, if one scheme achieves the same
precision than another scheme with smaller inverse cost, then we can say that
the former is more efficient than the later.

It is known that, for sufficiently small step-sizes $\tau$, the method $S(\tau)$
integrates exactly (up to exponentially small errors that are below machine
accuracy) a modified Hamiltonian system that is close to the original one.
Measuring the maximum variation of the energy along a given orbit will gives us
a good idea of how close is that modified Hamiltonian to the original
Hamiltonian.

Motivated by that, in all our numerical test, we measure the relative precision
of a given scheme applied with a given step-size $\tau$ by computing the maximum
variation  ($E_i =\max\{|H(t_0) - H(t)|\}$) of the energy along a given
numerical orbit obtained over $10^5$ steps of the method (with the same initial
conditions at the initial time $t_0$) and plot the $E_i$ versus the inverse cost
$\tau/s$. To fix criteria we will always consider step-sizes of the form:
$\tau_i = 1/2^i$ for $i=0,\dots,N$.

We are interested in very precise integrations of the Solar System, hence the
main goal is to determine for each scheme the maximum step-size ($\tau_i$) 
required to have an error in the energy variation up to machine accuracy. 

Through the paper we consider three test models that we believe illustrate  
different particularities of the Solar System and can be extrapolated to
other planetary systems. These are: a) the motion of the four inner planets
(Mercury to Mars); b) the motion of the four outer planets (Jupiter to Neptune)
and c) the motion of the eight planets on the Solar System (Mercury to Neptune).
The initial conditions and mass parameters have
been taken from the JPL Solar System ephemerides DEA405  
(\verb+http://ssd.jpl.nasa.gov/+).

Table~\ref{TPertSize} shows estimates on the size of the perturbation for
these three examples for both set of coordinates Jacobi and Heliocentric. To
estimate the size of the perturbation we have integrated each system over 100
years and computed the maximum values for $|H_{I}|$ and $|H_{K}|$ along
this integration. Here {\tt HKep} represents the size of the Keplerian part and
{\tt H1max} the size of the perturbation part and the estimated size of the
perturbation is given by $\eps = {\tt H1max}/{\tt HKep}$.

\begin{table}[htb]
\caption{
Size of the perturbation in Jacobi  and Heliocentric  coordinates
for the three test examples considered in this work: 4 inner planets (Mercury to Mars), first line;
4 outer planets (Jupiter to Neptune), middle line; 8 planets on the Solar system
(Mercury to Neptune), third line. 
}\label{TPertSize}
{\small 
\begin{center}
   \begin{tabular}{cc}
	\hline\noalign{\smallskip}
     {\tt Jacobi Coord.} & {\tt Heliocentric Coord.}\\
	\noalign{\smallskip}\hline \noalign{\smallskip}
 
    \begin{tabular}{ccc} 
     {\tt HKep} & {\tt H1max} & $\eps$ \\ 
    \noalign{\smallskip}\hline\noalign{\smallskip}
	
	 {\tt 1.3945E-04} & {\tt 6.3342E-10} & {\tt 4.5420E-06} \\
     {\tt 4.2924E-03} & {\tt 8.7162E-07} & {\tt 2.0306E-04} \\
     {\tt 4.4319E-03} & {\tt 8.7158E-07} & {\tt 1.9666E-04}
	\end{tabular}
	&
	\begin{tabular}{ccc}
       {\tt HKep} & {\tt H1max} & $\eps$ \\ 
    \noalign{\smallskip}\hline\noalign{\smallskip}
	  {\tt 1.3945E-04} & {\tt 9.1652E-10} & {\tt  6.5720E-06} \\
	  {\tt 4.2920E-03} & {\tt 2.7184E-06} & {\tt  6.3336E-04} \\
	  {\tt 4.4314E-03} & {\tt 2.8042E-06} & {\tt  6.3281E-04} \\
    \end{tabular} \\
	\noalign{\smallskip}\hline
\end{tabular}

\end{center}
}
\end{table}

We note that all the simulations in this article have been done using an extended real
arithmetics and that we use the compensated summation during the intermediate
evaluation of $\exp(a_i\tau A)$ and $\exp(b_i\tau B)$ (see Appendix~\ref{Apndx_CS}).


\section{Splitting Symplectic Integrators for Jacobi Coordinates}
\label{SymIntTST}

In Section~\ref{NBP_coord} we have seen that with an appropriate change of
variables we can rewrite the Hamiltonian of the N-body planetary system as
$H_{K} + H_{I}$ where $|H_I| \ll |H_K|$. Hence, the system falls into
the class of Hamiltonian that can be expressed as
\begin{equation}
H = H_{A} + \eps H_{B}, \label{EqInt}
\end{equation}
with $|\eps| \ll 1$. We can take advantage of this to build efficient
high-order splitting symplectic integrators~\citep{McLa9502,LaRo01}. In this
section we summarise the main ideas behind these methods and review some of the
most relevant schemes.

Using the Lie formalism the formal solution of Eq.~(\ref{EqInt}) is:
\begin{equation}
z(\tau) = \exp[\tau(A + \eps B)] z(\tau_0),
\label{SolHamiPert}
\end{equation}
where to simplify notation  we use $A \equiv \{H_A, \cdot\}=L_{H_A}$, $B \equiv \{H_B,
\cdot\}=L_{H_B}$. We recall that $H_A$ and $H_B$ are integrable, hence we can
compute explicitly $\exp(\tau A)$ and $\exp(\tau \varepsilon B)$.
To have a splitting symplectic integrator of order $r$, we need to find the
coefficients $a_i$, $b_i$ such that
\begin{equation}
\mathcal{S}_r(\tau) = \displaystyle\prod_{i=1}^{s}
	\exp(a_i\tau A) \exp(\varepsilon b_i\tau B),
\label{SplitMethDef}
\end{equation}
satisfies $|\mathcal{S}_r(\tau) - \exp[\tau(A + \varepsilon B)]| = \mathcal{O}(\tau^{r+1})$. The
Baker--Campbell--Hausdorff (BCH) theorem ensures us that $\mathcal{S}_r(\tau) =
\exp(\tau \mathcal{H})$, where $\mathcal{H}$ is also a Hamiltonian system and belongs to the free
Lie algebra generated by $A$ and $B$, $\mathcal{L}(A,B)$. Moreover, we can express $\mathcal{H}$ as a double
asymptotic series in $\tau$ and $\eps$:
\begin{equation}
\begin{array}{rcl}
\tau \mathcal{H} &=& \ \tau p_{1,0} A \ +\ \eps \tau p_{1,1} B \ +\ \eps \tau^2 p_{2,1} [A,B]
   + \eps \tau^3 p_{3,1} [[A,B],A]
\\ \\
	& + & \ \eps^2 \tau^3 p_{3,2} [[A,B],B] \ +\ \eps \tau^4 p_{4,1} [[[A,B],A],A]
\\ \\
   & + & \ \eps^2 \tau^4 p_{4,2}[[[A,B],B],A]
	\ +\ \eps^3 \tau^4 p_{4,3} [[[A,B],B],B] \ + \
	\dots,
\end{array} \label{ExpComp01}
\end{equation}
where $p_{i,j}$ are polynomials in $a_k$ and $b_k$. 

To have a symplectic scheme of order $r$ we need:
\begin{eqnarray*}
 p_{1,0} &=& a_1 \ +\  a_2 \ +\  \cdots \ +\  a_s = 1, \\ 
 p_{1,1} &=& b_1 \ +\  b_2 \ +\  \cdots \ +\  b_s = 1, \\
 p_{i,j} &=& 0, \ \forall i,j  \le r.
\end{eqnarray*}

The scheme $\mathcal{S}_r(\tau)$ is symmetric if it verifies
$\mathcal{S}^{-1}_r(\tau) = \mathcal{S}_r(-\tau)$, in which case
all the even terms in $\tau$
in Eq.~\ref{ExpComp01} vanish, having less conditions to satisfy for a scheme
of a given order, $r$, enabling us to find high-order schemes at lower computational
cost. There are two different types of symmetric compositions (Eq. \ref{SplitMethDef}): 
one in which the first and last exponentials correspond to the $A$ part (and thus called
$\mathcal{ABA}$ composition),
\begin{equation}   \label{aba.comp}
\mathcal{ABA}: \quad \e^{a_1 \tau A} \, \e^{\varepsilon b_1 \tau B} \, \e^{a_2 \tau A} \, \cdots  \,
     \e^{a_2 \tau A} \, \e^{\varepsilon b_1 \tau B} \, \e^{a_1 \tau A}
\end{equation}
and the other in which the role of $\exp(\tau A)$ and $\exp(\varepsilon \tau B)$ is
interchanged ($\mathcal{BAB}$ composition):
\begin{equation}   \label{bab.comp}
\mathcal{BAB}: \quad  \e^{\varepsilon b_1 \tau B} \, \e^{a_1 \tau A} \, \e^{\varepsilon b_2 \tau B} \, \cdots  \,
     \e^{\varepsilon b_2 \tau B}  \, \e^{a_1 \tau A} \, \e^{\varepsilon b_1 \tau B}.
\end{equation}
All the integration schemes that we present in this paper correspond to the $\mathcal{ABA}$ class.
For the experiments carried out, we have not found substantial differences in the efficiency
with respect to methods in the $\mathcal{BAB}$ class.

Notice that for symmetric methods, the last exponential at one step can be concatenated with the first
one at the next integration step when the method is iterated, so the number of exponentials
$\exp(\tau A)$ and $\exp(\varepsilon \tau B)$ per step is $s$, the number of stages.


It is clear that in many cases $|\eps| \ll \tau$ (or at least $\eps \approx
\tau$). So we can have high-order schemes by only killing the error terms with
small powers of $\eps$, and save computational cost by decreasing the number of
stages of the method.

Depending of the nature of the problem we can try to find the appropriate terms
in $\eps^i \tau^p$ that must vanish in order to have an optimal performance.
For example, if we consider a method such that the coefficients $a_i, b_i$
satisfy $p_{1,0} = p_{1,1} = 1$ and $p_{2,1} = p_{3,1} = p_{4,1} = 0$, then, $$
|\mathcal{H} - (A + \eps B)| = \mathcal{O} (\eps \tau^4 + \eps^2 \tau^2),$$ but as $|\eps|
\ll \tau$ this method is of effective order $4$.  In a more general context we
will have methods such that,
\begin{equation}
|\mathcal{H} - (A + \eps B)| =
\mathcal{O} (\eps \tau^{s_1} + \eps^2 \tau^{s_2} +
	\eps^3 \tau^{s_3} + \dots + \eps^m \tau^{s_m}).
\end{equation}
We remark that $s_1$ is the error of consistency for the scheme, i.e. is the
error behaviour in the limit case $\eps \rightarrow 0$. Nevertheless, in
many cases for small step-sizes the method can behave as one of order $s_2$.
In what follows we will refer to the  generalised order of a method in terms of the
order in powers of $\eps$. Hence, we will say that a method has
order $(s_1,s_2)$ if $|\mathcal{H} - (A + \eps B)| = \mathcal{O} (\eps \tau^{s_1} +
\eps^2 \tau^{s_2})$. In terms of the local error, we have
$|\mathcal{S}(\tau) - \exp[\tau(A + \varepsilon B)]| = \mathcal{O}(\eps \tau^{s_1+1} +
\eps^2 \tau^{s_2+1})$.

\subsection{$\mathcal{ABA}$ schemes of generalised order $(2n,2)$}
\label{SABAn}

\cite{McLa9502} noted that as $|\eps| \ll \tau$ we can have
high-order methods by only killing the terms in $\eps\tau^k$. Independently
\cite{Cham00,LaRo01} dealt with this problem following similar ideas, \citep{LaRo01}
providing an explicit computation of the coefficients of the remainder for all order $k$. One of
the main advantages of only killing the terms in $\eps\tau^k$ is that we are
sure that all the coefficients $a_i, b_i$ will be positive. As a consequence
the coefficients $a_i, b_i$ will be small and the numerical scheme will be
stable.

In Table~\ref{TAB_ABA_2s2} we summarise the coefficients for the different
$\mathcal{ABA}$ $(2n,2)$ schemes for $n = 1, \dots, 4$. For further details on
how to find the $a_i, b_i$ coefficients and the coefficients for $n\geq4$
see~\citep{McLa9502,LaRo01}. Since all the methods we consider are symmetric,
we only collect the necessary coefficients of each scheme. Thus,  $\mathcal{ABA}(8,2)$
corresponds to the composition
\[
  \e^{a_1 \tau A} \, \e^{b_1 \eps \tau B} \,  \e^{a_2 \tau A} \, \e^{b_2 \eps \tau B} \,
   \e^{a_3 \tau A} \, \e^{b_2 \eps \tau B} \,  \e^{a_2 \tau A} \, \e^{b_1 \eps \tau B} \,
    \e^{a_1 \tau A}.
\]
We will follow this convention throughout the text.

\begin{figure}[htb]
\centering
\includegraphics[width = 0.45\textwidth]{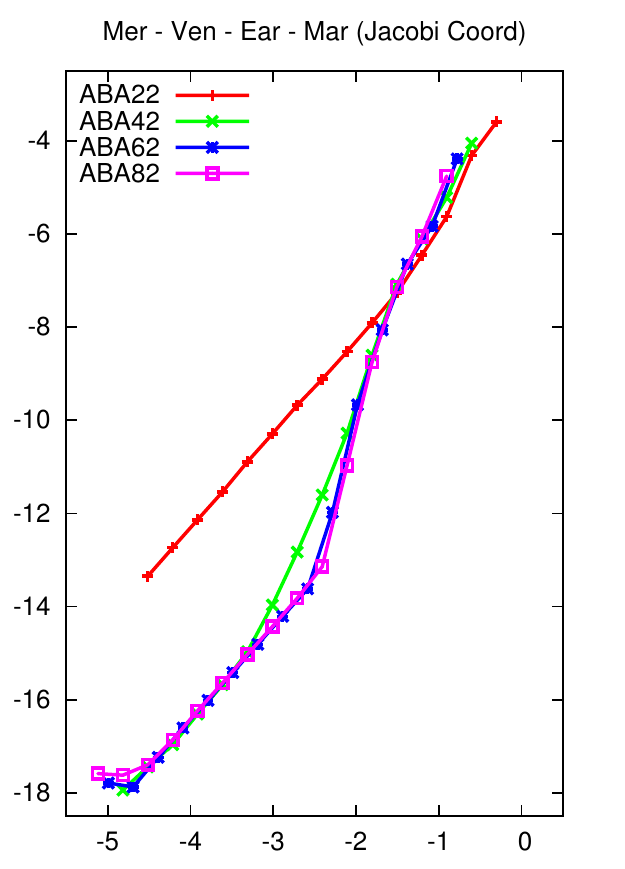}
\includegraphics[width = 0.45\textwidth]{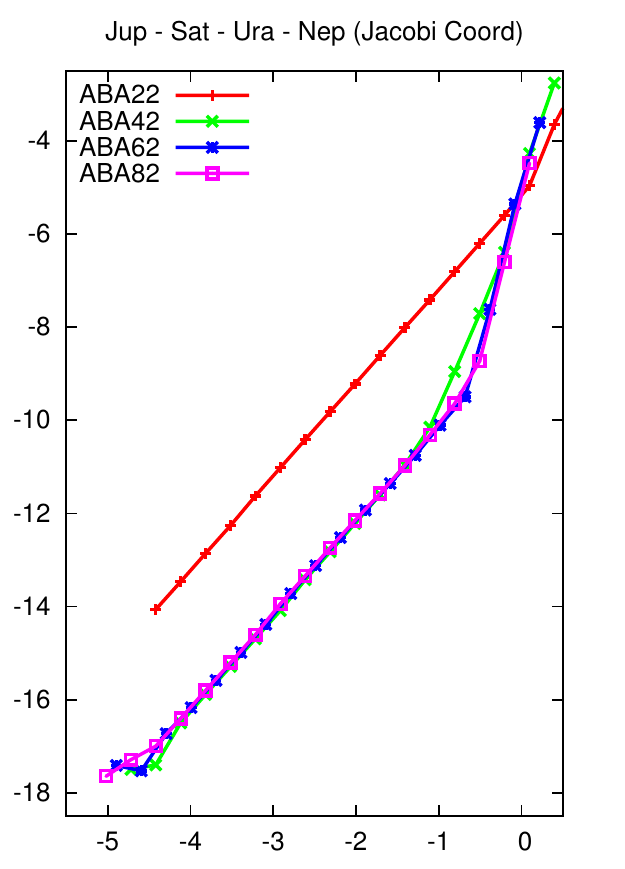}
\caption{
Comparison between the $\mathcal{ABA}(2n, 2)$ methods for $n = 1,2,3, 4$
applied to the 4 inner planets (left) and the 4 outer planets (right). The $x$-axis
represents the cost ($\tau/s$) and the $y$-axis is the maximum energy variation over
one integration with constant step-size $\tau$. Here $s$ is the number of stages.}\label{FigSABAn}
\end{figure}

In Figure~\ref{FigSABAn} we compare the performance of the $\mathcal{ABA}(2n,2)$
for $n={1,2,3,4}$ for the 4 inner planets (left) and the 4 outer planets (right).
The $x$-axis corresponds to the cost of the scheme $(\tau/s)$ and the $y$-axis
corresponds to the maximum energy variation for one integration at constant
step-size $\tau$. \cite{LaRo01} already saw that the optimal schemes for this
problem where those of orders $(6,2)$ and $(8,2)$ (i.e. $\mathcal{SABA}_3$ and
$\mathcal{SABA}_4$ following their notation).

\begin{table}[htb]
\caption{
Coefficients for the $\mathcal{ABA}(2n, 2)$ methods for $n = 1, \dots, 4$
\citep{LaRo01}.
} \label{TAB_ABA_2s2}
\centering
\begin{tabular}{cccl}
\hline\noalign{\smallskip}
{\tt id} & {\tt order} & {\tt stages}& $a_i, b_i$ \\
\noalign{\smallskip}\hline\noalign{\smallskip}
{\tt ABA22} & $(2,2)$ & 1 &
{\footnotesize \begin{tabular}{rcl}
	$a_1$ &=& $1/2$  \\
	$b_1$ &=& $1  $  \\
\end{tabular} } \\
\noalign{\smallskip}\hline\noalign{\smallskip}
{\tt ABA42} & $(4,2)$ & 2 &
{\footnotesize \begin{tabular}{rcl}
	$a_1$ &=& $1/2 - \sqrt{3}/6$ \\
	$a_2$ &=& $\sqrt{3/3}$ \\
	$b_1$ &=& $1/2$
\end{tabular} } \\
\noalign{\smallskip}\hline\noalign{\smallskip}
{\tt ABA62} & $(6,2)$ & 3 &
{\footnotesize	\begin{tabular}{rcl}
   $a_1$ &=& $1/2 - \sqrt{15}/10$ \\
	$a_2$ &=& $\sqrt{15}/10$ \\
   $b_1$ &=& $5/18$ \\
	$b_2$ &=& $4/9$
\end{tabular} }\\
\noalign{\smallskip}\hline\noalign{\smallskip}
{\tt ABA82} & $(8,2)$ & 4 &
{\footnotesize	\begin{tabular}{rcl}
	$a_1$ &=& $1/2 -\sqrt{525+70\sqrt{30}}/70$ \\
	$a_2$ &=& $\left(\sqrt{525+70\sqrt{30}}-\sqrt{525-70\sqrt{30}}\right)/70$ \\
	$a_3$ &=& $\sqrt{525-70\sqrt{30}}/35$ \\
	$b_1$ &=& $1/4 - \sqrt{30}/72$ \\
	$b_2$ &=& $1/4 + \sqrt{30}/72$
	\end{tabular} }\\
\noalign{\smallskip}\hline\noalign{\smallskip}
\end{tabular}
\end{table}

The error on the Hamiltonian approximation of these schemes is
$\mathcal{O}(\eps\tau^{2n} + \eps^2\tau^2)$. In Figure~\ref{FigSABAn} we can see
how the error in energy decreases in $\tau$ with slope $2n$ for large steps-sizes and
slope $2$ for smaller steps-sizes. We also see how for small step-sizes
there is no difference between the cost of the $\mathcal{ABA}62$ and
$\mathcal{ABA}82$ schemes. In order to improve their performance we need to
kill the term in $\eps^2\tau^2$ rather than those of order $\eps\tau^{2k}$
for $k>4$, which are the limiting factor of these schemes.

\subsection{$\mathcal{ABA}$ schemes of order $(2n,4)$}
\label{ABA84}




In this section we will describe three different procedures to 
cancel the dominant
term  $\eps^2\tau^2$ in order to get methods of generalized order $(2n,4)$,
and discuss their
performance for the different test models described in
Section~\ref{TEST_MODEL}.

\subsubsection{The corrector term ($\mathcal{SC}$)} \label{SABACn}


Since in Jacobi coordinates $A$ is quadratic in $p$ and $B$ depends only of $q$,
then it follows that the term $[[A,B],B]$ depends only on $q$ and thus
$\exp( \tau^3 \eps^2 [[A,B],B])$ can be easily computed.
\cite{LaRo01} noticed that it is possible
to incorporate this term into the previous compositions with a conveniently chosen constant
so as to cancel the term of order $\eps^2 \tau^2$ in the asymptotic expansion
Eq.~\ref{ExpComp01}.
We note that this corrector scheme is
different than the one introduced by \cite{WiHoTo96} where the corrector added
at the beginning and at the end of each step-size is a change of variables.


Thus, let $\mathcal{S}_n(\tau)$ be one of the symplectic $\mathcal{ABA}$ schemes of
order $(2n,2)$ described in Section~\ref{SABAn}. We can get rid of the term
in $\eps^2 \tau^2$ by considering
\begin{equation}
\mathcal{SC}_n(\tau) = \exp\left(-\tau^3\eps^2 \frac{c}{2}[[A,B],B]\right)\
	\mathcal{S}_n(\tau)\ \exp\left(-\tau^3\eps^2 \frac{c}{2}[[A,B],B]\right),
\label{SchemeCorr}
\end{equation}
with the appropriate choice of the constant $c$. In Table~\ref{TAB_ABAC_2s2} we
show the value for the coefficient $c$ for each the four $\mathcal{ABA}(2n,2)$
schemes described before. For further details see~\citep{LaRo01}. Notice that
$\mathcal{SC}_n$ corresponds to integrating $\log(\mathcal{S}_n(\tau))~-\tau^3\eps^2 c L_{\{\{A,B\}, B\}}$
using the leapfrog scheme.

So using Eq.~\ref{SchemeCorr} with any of the $\mathcal{ABA}$ $(2n,2)$ scheme
in Section~\ref{SABAn} we obtain a new integrating scheme of order $(2n, 4)$ with
no negative intermediate step.

\begin{table}[htb]
\caption{
Coefficients $c$ for the corrector term applied to the $\mathcal{ABA}(2n,2)$
schemes in Table~\ref{TAB_ABA_2s2} \citep{LaRo01}.
}\label{TAB_ABAC_2s2}
\centering
\begin{tabular}{cl} 
\hline\noalign{\smallskip}
{\tt order} & $c$ \\ 
\noalign{\smallskip}\hline\noalign{\smallskip}
$1$ & {\footnotesize $1/12$ } \\ 
$2$ & {\footnotesize $(2 - \sqrt{3})/24$ } \\ 
$3$ & {\footnotesize $(54 - 13\sqrt{15})/648$ } \\ 
$4$ & {\footnotesize $0.003396775048208601331532157783492144$ } \\ 
\noalign{\smallskip}\hline
\end{tabular}
\end{table}

\subsubsection{The composition scheme ($\mathcal{S}2^m)$}
\label{SABAn_comp}

\cite{Yosh90} and \cite{Suzu90} independently came up with the same idea to find
a symmetric scheme of order $2k+2$ from one of order $2k$. They both noticed that if
$\mathcal{S}(\tau)$ is a scheme of order $2k$, then:
\begin{equation}
  \mathcal S2k(\tau) =
  \mathcal{S}(x_0\tau) \mathcal{S}(x_1\tau) \mathcal{S}(x_0\tau),
\end{equation}
is a scheme of order $2k+2$ for an appropriate choice of the constant
coefficients $x_0,x_1$. One can check that $x_0,x_1$ must satisfy $2x_0 + x_1 =
1$ and $2x_0^{2k+1} + x_1^{2k+1} = 0$.  Notice that the second condition is
used to cancel all the terms of order $2k$, while the first one is only 
for consistency.

 \cite{LaRo01} used this idea to turn any of the $\mathcal{ABA}$
 schemes of
 order $(2n,2)$ into one of order $(2n,4)$. If $\mathcal S(\tau)$ is a
 symmetric $\mathcal{ABA}$ scheme of order (2n,2) then the
 composition:
 \begin{equation}
    \mathcal S2^{m}(\tau) =
    \mathcal S^{m}(y_0\tau) \mathcal S(y_1 \tau) \mathcal
 S^m(y_0\tau).
 \end{equation}
 is a symmetric method of order $(2n,4)$ if $y_0,y_1$ satisfy $2my_0 +
 y_1 =
 1$ and $2my_0^{3} + y_1^{3} = 0$ so, ($y_0=1/(2m-(2m)^{1/3}), \
 y_1=-(2m)^{1/3}/(2m-(2m)^{1/3})$).
 
%
%

We have done several test to determine the optimal value of $m$, and our test show
that this one is given by $m = 2$. These results are consistent with
those of~\citep{Suzu90,McLa02} who did a similar study in a more general
framework. The main advantage of this scheme is that we can use it for both Heliocentric
and Jacobi coordinates.


\subsubsection{McLachlan extra stage scheme ($\mathcal{ABA}84$)}
\label{McLahan84}

\cite{McLa9502} studied the possibility of adding an extra stage to the
$\mathcal{ABA}(2n,2)$ schemes to get rid of the $\eps^2\tau^2$ term. To add
an extra stage derives into having an extra pair of coefficients $a_i,b_i$ and an
extra algebraic equation to satisfy. All the coefficients will no longer
be positive~\citep{Suzu91}.
In general, if the coefficients are not very large, these methods are stable. The
coefficients for the $\mathcal{ABA}$  method of generalised order $(8,4)$
provided by \cite{McLa9502} are
summarised in Table~\ref{TAB_ABA_2s4}.

\begin{table}[htb]
\caption{
Coefficients for the $\mathcal{ABA}$ method of order
$(8, 4)$ found by \citep{McLa9502}.
}\label{TAB_ABA_2s4}
\centering
\begin{tabular}{cccl}
\hline\noalign{\smallskip}
{\tt id} & {\tt order} & {\tt stages} & $a_i, b_i$ \\
\noalign{\smallskip}\hline\noalign{\smallskip}
{\tt ABA84} & $(8,4)$ & 5 &
{\footnotesize	\begin{tabular}{rcl}
$a_1$ &=& {\tt \ 0.07534696026989288841652780368} \\
$a_2$ &=& {\tt \ 0.51791685468825678230077397850 } \\
$a_3$ &=& {\tt  -0.09326381495814967071730178218} \\
$b_1$ &=& {\tt \ 0.19022593937367661924523076274} \\
$b_2$ &=& {\tt \ 0.84652407044352625705508054465} \\
$b_3$ &=& {\tt  -1.07350001963440575260062261477}
	\end{tabular} }\\
\noalign{\smallskip}\hline
\end{tabular}
\end{table}

\subsubsection{Results}

\begin{figure}[htb]
\centering
\includegraphics[width = .32\textwidth]{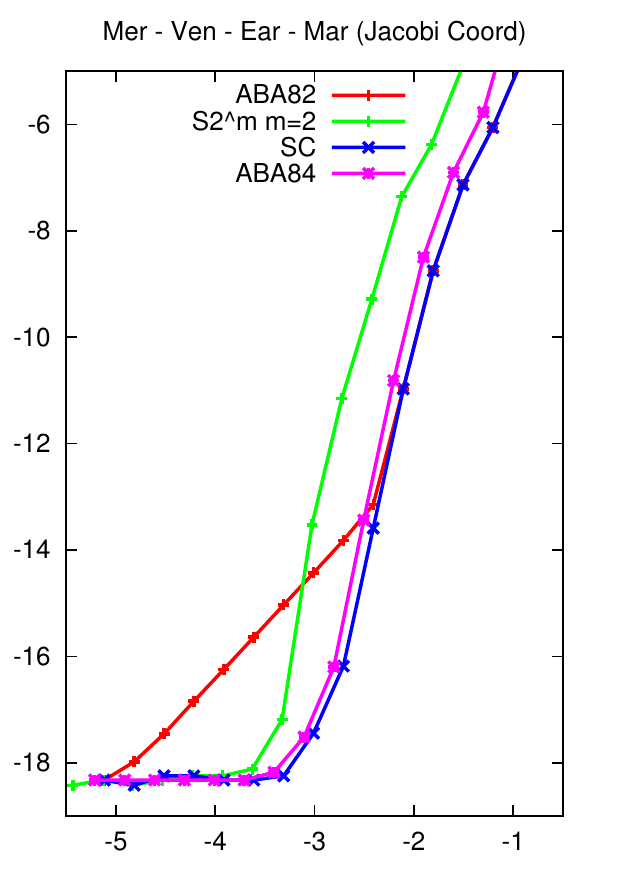}
\includegraphics[width = .32\textwidth]{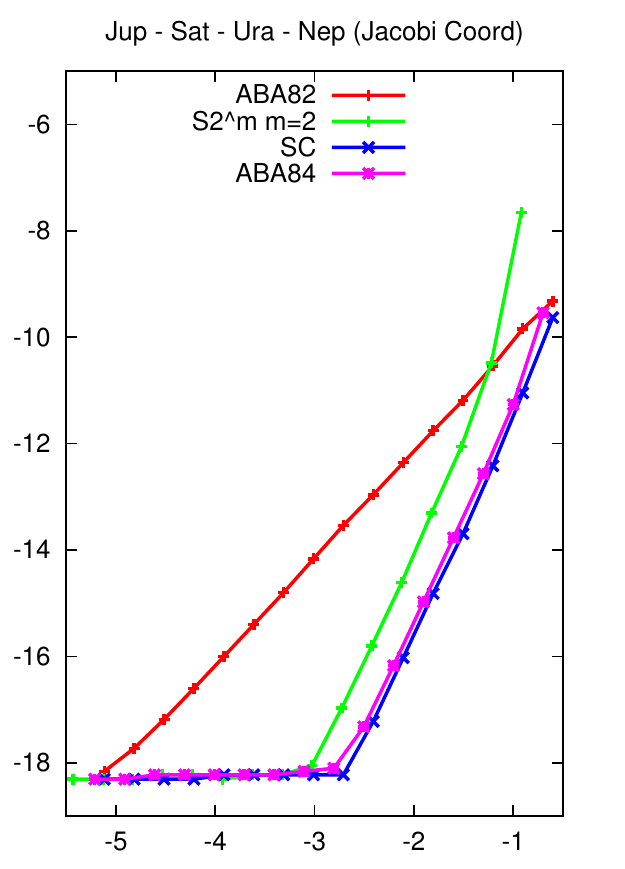}
\includegraphics[width = .32\textwidth]{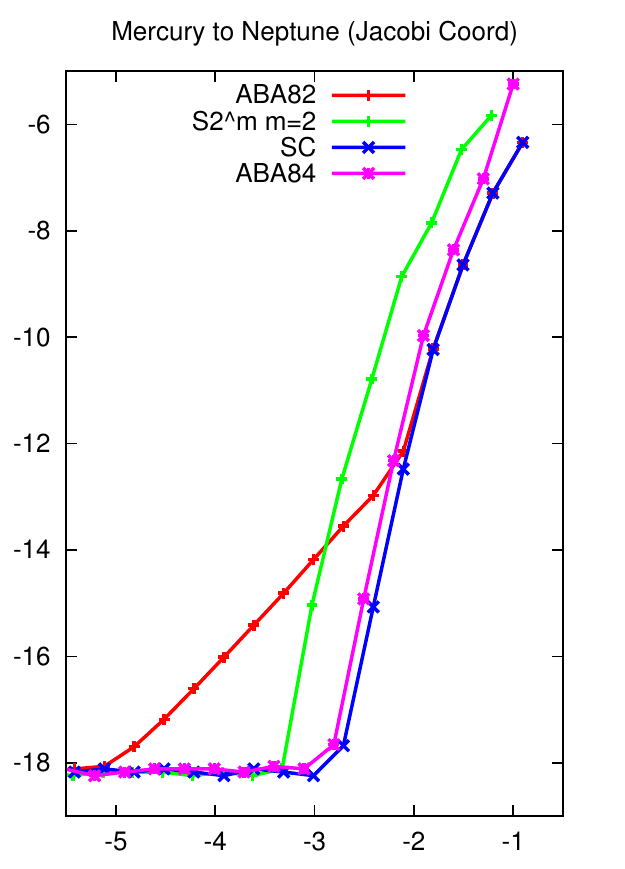}
\caption{Comparison between the different schemes to kill the
the $\eps^2\tau^2$ terms in the $\mathcal{ABA}82$ scheme. From
left to right: the 4 inner planets, the 4 outer planets and the whole Solar
System. The $x$-axis represents the cost $(\tau/s)$ and the $y$-axis the
maximum energy variation for one integration with constant step-size $\tau$.
}\label{FigABA84}
\end{figure}

In Figure~\ref{FigABA84} we compare the performance of these three different
approaches to build methods of generalised order (8,4)
against the $\mathcal{ABA}82$
scheme. In the plots we show the cost $(\tau/s)$ vs the maximum energy
variation for the three test models: the 4 inner planets (left), the 4 outer
planets (middle) and the 8 planets in the Solar System (right).

As we can see, the three different schemes improve considerably the performance
of the $\mathcal{ABA}82$ (red line). In all cases the corrector scheme
$\mathcal{SC}$ (blue line) and the McLachlan $\mathcal{ABA}84$ (purple line)
show a similar quantitative behaviour. The difference between them is the cost
of the extra stage in $\mathcal{ABA}84$, as we are assuming that the corrector is
completely free. We note that this is not entirely true if the number of bodies is
large ($n\geq4$).
On the other hand, the composition methods, $\mathcal{S2}^m$ (green line),
improves the performance with respect to the $\mathcal{ABA}82$ (red line) but
is much more expensive than the other two schemes.

\subsection{$\mathcal{ABA}$ schemes with generalised order $(s_1,s_2, \dots)$}
\label{BlanesSplit}

In the previous section we have seen that adding an extra stage to cancel the term
of order $\eps^2\tau^2$ gives good results. We can extend this idea and add
more stages to kill the error terms accounting to the main limiting
factor for each problem~\citep{Bla12}. This translates into adding more constraints on the
coefficients. As long as the increase in the
computational cost is less than the gain in accuracy these methods will be
competitive.
In Figure~\ref{FigABA84} we see that the dominant error term for the
$\mathcal{ABA}84$ varies between the different test models. Notice that for the
outer planetary system the scheme behaves as one of order $4$, so the
dominant term is $\eps^2\tau^4$. On the other hand, for the inner planetary
system, the method behaves as one of order 8, now the dominant term is
$\eps\tau^8$.

Hence, to improve the performance of the McLachlan's $\mathcal{ABA}84$ we need
to kill different error terms depending on the problem. For the inner planets,
a method of order $(10,4)$ should perform better than one of order $(8,6)$.
While for the outer planets a method of order $(8,6)$ should give better
results than one of order $(10,4)$.

In~\cite{Bla12} we find details on how to solve the algebraic equations and
find the set of coefficients $a_i,b_i$ that provide $\mathcal{ABA}$
 schemes for a given arbitrary order $(s_1,s_2, \dots)$. We must mention
that there is no unique combination of coefficients $a_i,b_i$ for a given order.
From all the possible solutions we have selected those that give a better
approximation and whose coefficients $a_i,b_i$ are small.
In Table~\ref{BlanesMethod} we summarise the coefficients for three $\mathcal{ABA}$
schemes: one of order $(10,4)$; one of order $(8,6,4)$ and one of order $(10,6,4)$.

\begin{table}[htb]
\caption{
Coefficients for $\mathcal{ABA}$ symmetric splitting methods of
orders $(10,4)$, $(8,6,4)$ and $(10,6,4)$ \citep{Bla12}.
}\label{BlanesMethod}
\centering
\begin{tabular}{cccl}
\hline\noalign{\smallskip}
{\tt id} & {\tt order} & {\tt stages}& $a_i,  b_i$ \\
\noalign{\smallskip}\hline\noalign{\smallskip}
{\small {\tt ABA104}} & {\small $(10,4)$} & {\small 7} &
{\footnotesize \begin{tabular}{rcl}
$a_1$ &=& {\tt \ 0.047067100645972506129478876372} \\ 
$a_2$ &=& {\tt \ 0.184756935417088106924737619370} \\ 
$a_3$ &=& {\tt \ 0.282706005679836205324361656554} \\ 
$a_4$ &=& {\tt  -0.014530041742896818378578152296} \\ 
$b_1$ &=& {\tt \ 0.118881917368197019945350395085}\\ 
$b_2$ &=& {\tt \ 0.241050460551501565744166786590}\\ 
$b_3$ &=& {\tt  -0.273286666705323806054311398166}\\ 
$b_4$ &=& {\tt \ 0.826708577571250440729588432981}\\ 
 	\end{tabular} }\\
\noalign{\smallskip}\hline\noalign{\smallskip}
{\small {\tt ABA864}} & {\small $(8,6,4)$} & {\small 7} &
{\footnotesize \begin{tabular}{rcl}
$a_1$ &=& {\tt \ 0.071133426498223117777938730006} \\ 
$a_2$ &=& {\tt \ 0.241153427956640098736487795326} \\ 
$a_3$ &=& {\tt \ 0.521411761772814789212136078067} \\ 
$a_4$ &=& {\tt  -0.333698616227678005726562603400} \\ 
$b_1$ &=& {\tt \ 0.183083687472197221961703757166} \\ 
$b_2$ &=& {\tt \ 0.310782859898574869507522291054} \\ 
$b_3$ &=& {\tt  -0.026564618511958800697212137916} \\ 
$b_4$ &=& {\tt \ 0.065396142282373418455972179391} \\ 
   \end{tabular} }\\
\noalign{\smallskip}\hline\noalign{\smallskip}
{\small {\tt ABA1064}} & {\small $(10,6,4)$} & {\small 8} &
{\footnotesize \begin{tabular}{rcl}
$a_1$ &=& {\tt \ 0.038094497422412195456975322308} \\ 
$a_2$ &=& {\tt \ 0.145298716116913749294020072660} \\ 
$a_3$ &=& {\tt \ 0.207627695725541250716205611324} \\ 
$a_4$ &=& {\tt \ 0.435909703651526159223154862401} \\ 
$a_5$ &=& {\tt  -0.653861225832786709380711737390} \\ 
$b_1$ &=& {\tt \ 0.095858880837075210610771503771} \\ 
$b_2$ &=& {\tt \ 0.204446153142998780680507783916} \\ 
$b_3$ &=& {\tt \ 0.217070347978991101714338592430} \\ 
$b_4$ &=& {\tt  -0.017375381959065093005617880118} \\ 
  \end{tabular} }\\
\noalign{\smallskip}\hline
\end{tabular}
\end{table}

\subsubsection{Results}

\begin{figure}[htb]
\centering
\includegraphics[width = .32\textwidth]{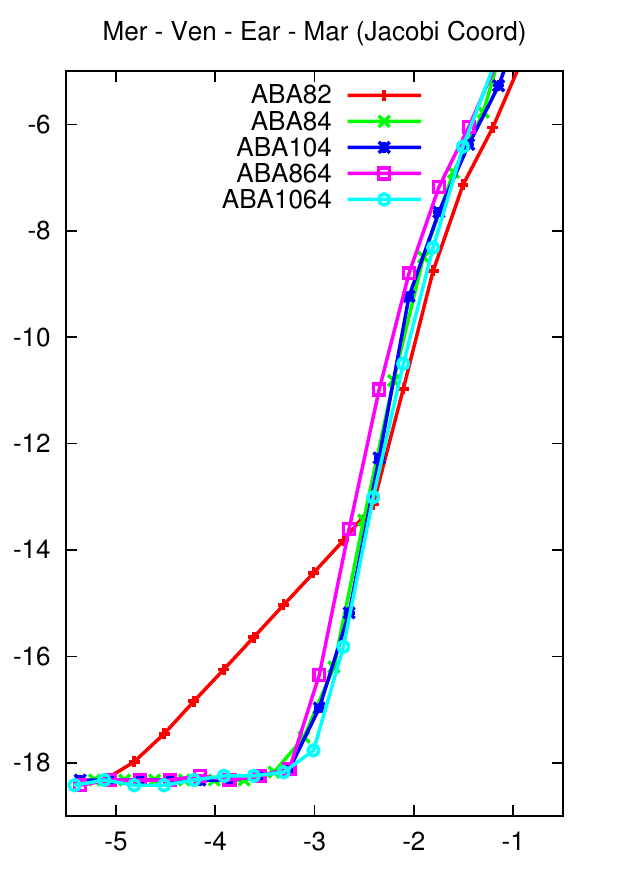}
\includegraphics[width = .32\textwidth]{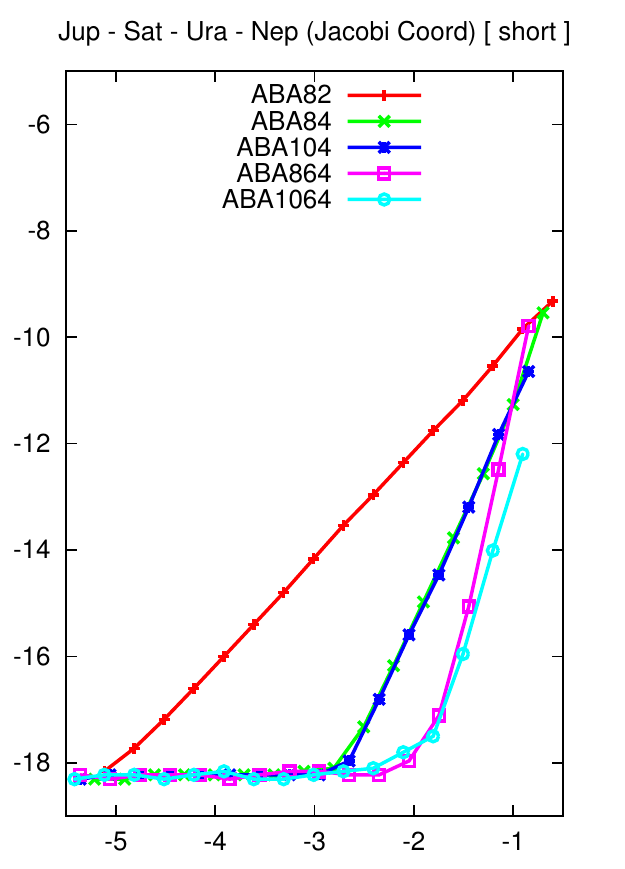}
\includegraphics[width = .32\textwidth]{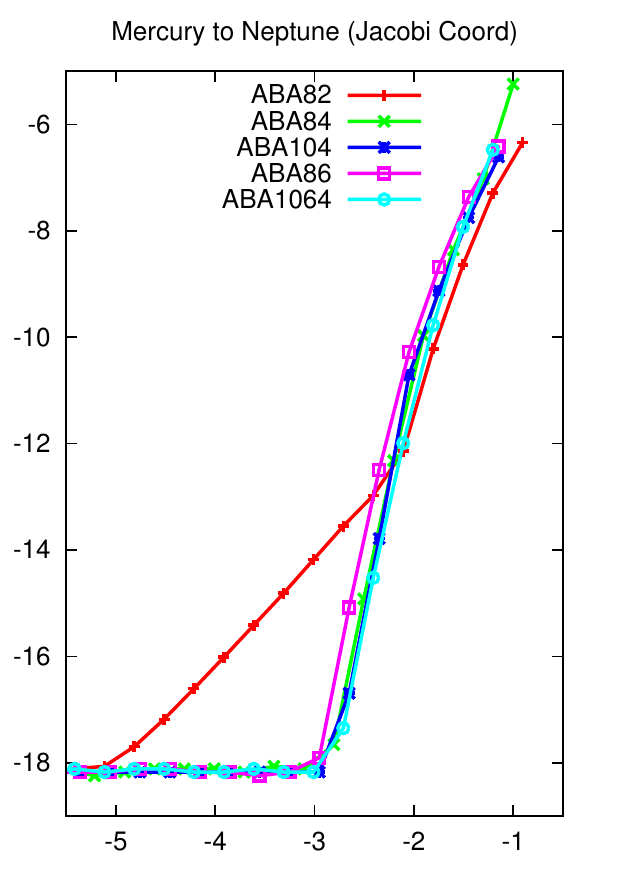}
\caption{Comparison between the $\mathcal{ABA}$ splitting schemes of arbitrary order
$(s_1,s_2,s_3)$ summarised in Table~\ref{BlanesMethod} against the $\mathcal{ABA}82$
and $\mathcal{ABA}84$. From left to right: the 4 inner planets, the 4 outer
planets and the whole Solar System. The $x$-axis represents the cost $(\tau/s)$ and
the $y$-axis the maximum energy variation for one integration with constant
step-size $\tau$.}\label{FigABAHigh}
\end{figure}

In Figure~\ref{FigABAHigh} we compare the performance of the three schemes
summarised in Table~\ref{BlanesMethod} against the $\mathcal{ABA}82$ and
$\mathcal{ABA}84$ for the three test models.

In the left-hand side of Figures~\ref{FigABAHigh} we have the results for the 4
inner planets. We recall that the dominant error term in the $\mathcal{ABA}84$
scheme was $\eps\tau^8$. Hence, a method of order $10$ in $\eps$ should perform
better than one of order $8$ in $\eps$. Nevertheless, as we can see there is
no significant gain in the performance of these schemes with respect to
$\mathcal{ABA}84$. Apparently, for these methods
the gain in precision is proportional
to the computational cost in this range of accuracy.

In the middle of Figure~\ref{FigABAHigh} we have the results for the 4 outer
planets. We recall that here the dominant term in the $\mathcal{ABA}84$ scheme
was $\eps^2\tau^4$, hence we expect the schemes of order $6$ in $\eps^2$ to be
better than the $\mathcal{ABA}84$. As we can see the $\mathcal{ABA}864$ and the
$\mathcal{ABA}1064$ schemes give better results that the $\mathcal{ABA}84$. In
both cases the optimal cost is around $10^{-2}$ vs an optimal cost of around 
$10^{-3}$ for the $\mathcal{ABA}84$ scheme.

The main difference between the inner planets and the outer planets is the size
of the perturbation. We recall that in Jacobi coordinates, for the inner
planets $\eps \approx 4.54\cdot 10^{-6}$, while for the outer planets $\eps
\approx 2.03\cdot10^{-4}$ (see Table~\ref{TPertSize}). The difference of about
2 orders of magnitude should explain the difference in the performance of the
different schemes, as the relevance of the terms $\eps^i\tau^{2k}$ in the error
approximation will vary depending on the size of $\eps$.

Taking this into account, one can be surprised by the performance of these
schemes when we consider the whole Solar System (Figure~\ref{FigABAHigh}
right). Here the size of the perturbation ($\eps \approx 1.96\cdot10^{-4}$) is
of the same order of magnitude as the case of the outer planets. But as we can
see in Figure~\ref{FigABAHigh} the schemes behave in the same way as the case
of the inner planets, where the terms of order $\eps\tau^8$ dominate those of
order $\eps^2\tau^4$. We think this is due to Mercury: its fast orbital period
and large eccentricity is limiting the performance of the methods. This
phenomena was already noticed by \cite{WiHoTo96} and re-discussed
by~\cite{Visw02}. This is why \cite{SaTre94} proposed to use independent
time-steps for each planet. They used the leapfrog scheme and adapted it to
take fractions of the given step-size for each planet, depending on their
orbital period. It is not trivial to extend these ideas using the higher order
schemes described in this section,
{and a second order method is not the appropriate
option to achieve round-off accuracy}.

\section{Splitting Symplectic Integrators for Heliocentric}\label{SymIntHelio}

We recall that all the tests done in Section~\ref{SymIntTST} have been done
using Jacobi coordinates. All these integrating schemes assume that the two
parts of the Hamiltonian $H_A, H_B$ are integrable. This is true for Jacobi
coordinates where:
\begin{equation}
H_{Jb}(q,p) = H_{K}(q,p) + H_{I}(q).
\end{equation}
where $H_{K}(q,p)$ is integrable (it is a sum of independent Kepler
problems) as well as $H_{I}(q)$ (it only depends on $q$). However, this is not true
for Heliocentric coordinates where:
\begin{equation}
H_{He}(q,p) = H_{K}(q,p) + H_{I}(q,p),
\end{equation}
and $H_{I}(q,p)$ is not integrable, which can be a problem if we want to apply
the splitting schemes presented in Section~\ref{SymIntTST}.

An option to deal with the non-integrability of $H_{I}(q,p)$ is to use another
numerical method to integrate $H_{I}(q,p)$ and compute the $\exp(b_i\tau B)$
up to machine accuracy. Unfortunately, this can drastically increase the
computational cost of the schemes.

We propose to use the fact that $H_{I}(q,p) = T_1(p) + U_1(q)$ splits naturally
into two parts, one depending on positions, the other in velocities. These two
parts are integrable on its own and small with respect to $H_{K}(q,p)$.  In a
general framework, the Hamiltonian splits as:
$$H = H_{A} + \eps (H_{B} + H_{C}),$$
where $H_{A}$, $H_{B}$ and $H_{C}$, are integrated when we consider them
separately. In the same way as in Section~\ref{SymIntTST}, we could try to find
appropriate coefficients $a_i,\ b_i,\ c_i$, such that
$$
\mathcal{S}(\tau) =
	\prod_{i = 1}^s \exp(a_i\tau A)\exp(\eps b_i\tau B)\exp(\eps c_i\tau C),
$$
approximates $\exp(\tau L_H)$. As before, to simplify notation 
$A\equiv\{H_A, \cdot\}$, $B\equiv\{H_B, \cdot\}$ and $C\equiv\{H_C, \cdot\}$. 
Then one has to deal with the Lie Algebra
generated by $A, B$ and $C$. The number of order conditions as
well as the complexity to solve them numerically to get the coefficients
$a_i,b_i,c_i$ grows extraordinarily with the order \cite{Bla12}.
A simple alternative way to proceed is to use the splitting symplectic schemes
in Section~\ref{SymIntTST}:
\begin{equation}
\mathcal{S}(\tau) =
   \prod_{i = 1}^n \exp(a_i\tau A)\exp(\eps b_i\tau (B+C)).
\label{SymHelio01}
\end{equation}
and approximate $\exp(\eps b_i\tau (B+C))$ with
\begin{equation}
	\exp(\eps b_i\tau (B+C)) \approx
	\exp(\eps \frac{b_i}{2}\tau C)\exp(\eps b_i\tau B)\exp(\eps \frac{b_i}{2}\tau C).
\label{eq:leaptric}
\end{equation}
Here we take $C$ as the Lie operator associated to $T_1(p)$ due to
its lower computational cost.

\begin{figure}[h!]
\centering
\includegraphics[width = 0.32\textwidth]{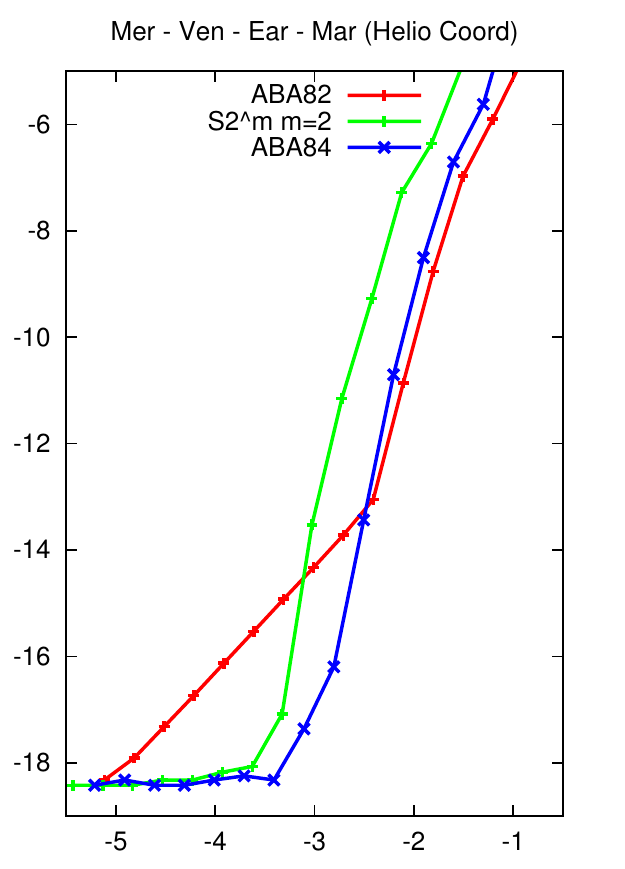}
\includegraphics[width = 0.32\textwidth]{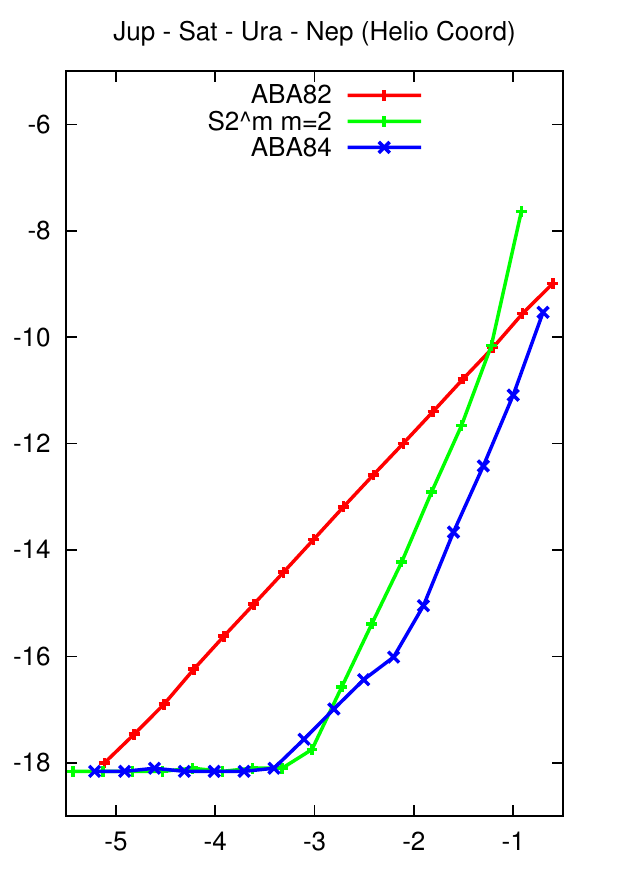}
\includegraphics[width = 0.32\textwidth]{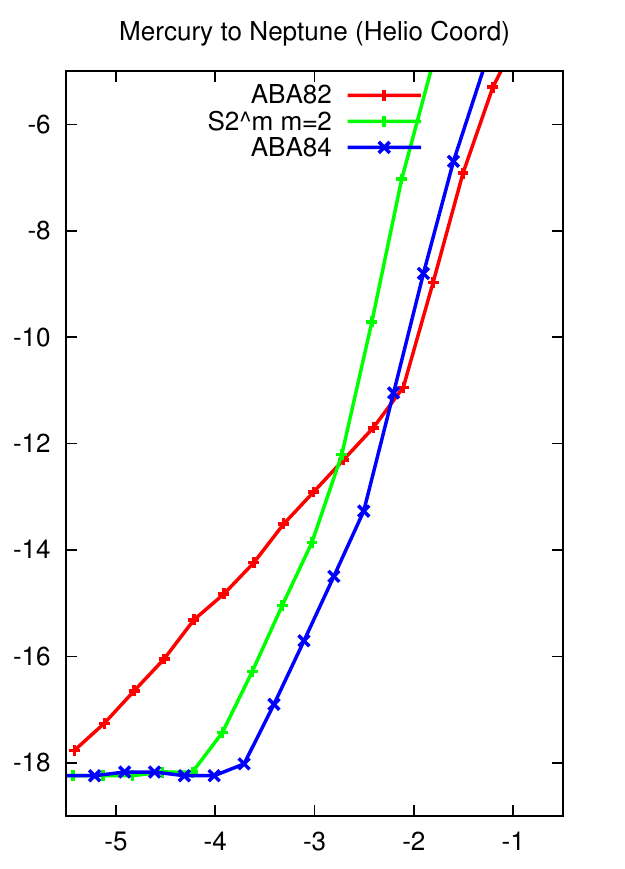}
\caption{
Comparison between the $\mathcal{ABA}82$, $\mathcal{ABA}84$ and $\mathcal{S}2^m$
schemes discussed in Section~\ref{ABA84} applied to Heliocentric coordinates.
From left to right: the 4 inner planets, the 4 outer planets and the whole Solar
System. The $x$-axis represents the cost ($\tau/s$) and the $y$-axis the maximum
energy variation for one integration with constant step-size $\tau$.
}\label{FigABAH8401}
\end{figure}

In general $H_B$ and $H_C$ do not commute ($\{H_B, H_C\} \neq 0$), so this
approximation adds an extra error contribution term, $\eps^3\tau^2$,
which will be negligible for small $\eps$.
In Figure~\ref{FigABAH8401} we see the result of taking the $\mathcal{ABA}82$,
the $\mathcal{ABA}84$ and the $\mathcal{S}2^m$ splitting schemes using
Eq.~\ref{eq:leaptric} to deal with Heliocentric coordinates. We can see that in
general the symplectic schemes have the same behaviour as with Jacobi
coordinates (Figure~\ref{FigABA84}).

We do see a difference in the case of the outer planets (Figure~\ref{FigABAH8401}
middle). Now the $\mathcal{ABA}84$ scheme behaves as one of order $2$ for small
step-sizes. This is due to the extra error term $\eps^3\tau^2$. We recall that
the main difference between the inner and the outer planets is the size
of the perturbation ($\eps$) which is smaller in the first case. Here the terms
of order $\eps^3\tau^2$ are negligible for the inner planets but not for the
outer planets.

Unfortunately, when we consider high-order symplectic schemes like the ones
presented in Sections~\ref{BlanesSplit} these extra error term will become
relevant, jeopardising the performance of these schemes.

\cite{Cham99,Wisd06} proposed to rewrite the Hamiltonian in Heliocentric
variables so that $H_B$ and $H_C$ commuted ($\{H_B,H_C\} = 0$), and then used
the fact that $\exp(\eps b_i\tau (B+C)) = \exp(\eps b_i\tau B)\exp(\eps b_i\tau
C)$. For further details see Appendix~\ref{Apndx_HvsH}.

\subsection{$\mathcal{ABAH}$ $(2n,4)$ specific for Heliocentric coordinates}
\label{ABAHn}

\begin{figure}[h!]
\centering
\includegraphics[width = 0.32\textwidth]{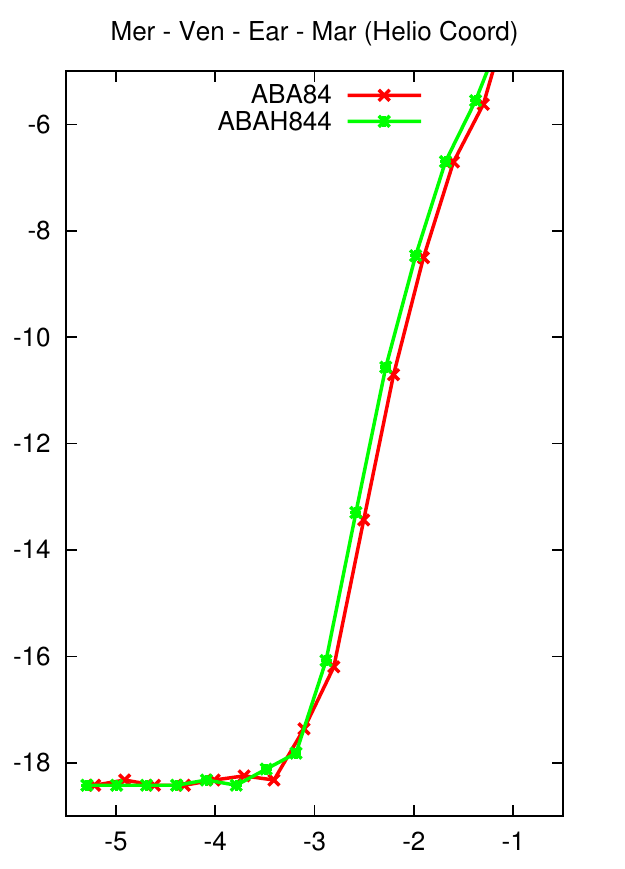}
\includegraphics[width = 0.32\textwidth]{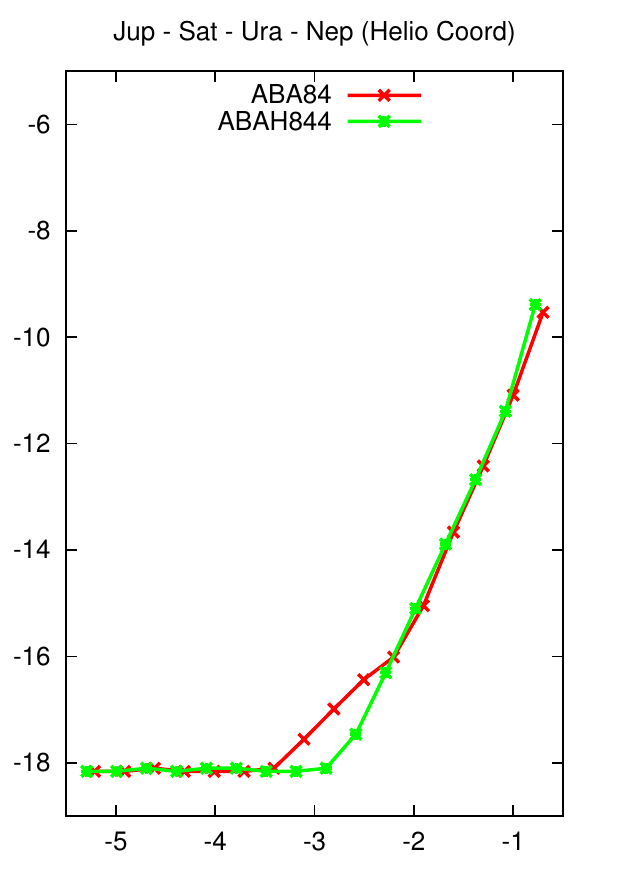}
\includegraphics[width = 0.32\textwidth]{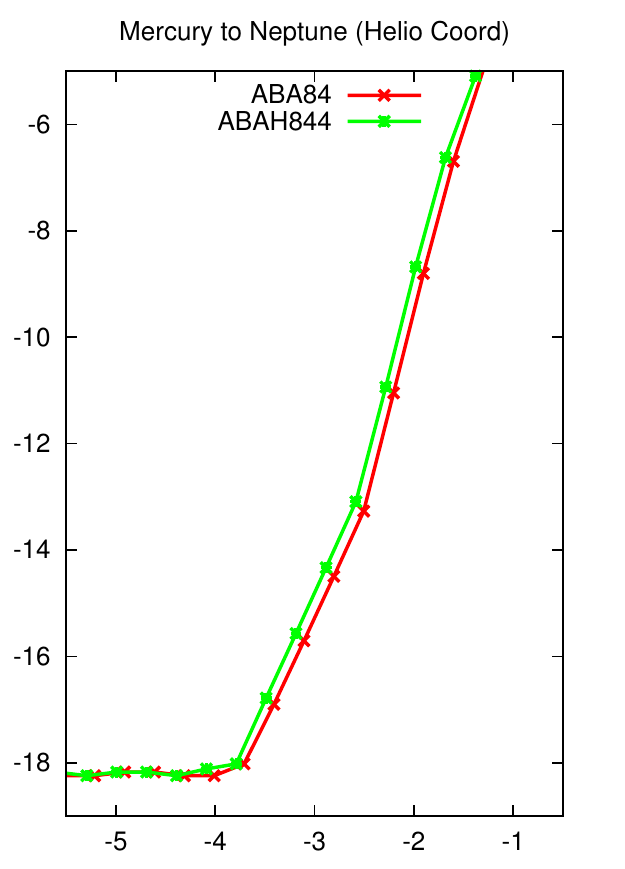}
\caption{
Comparison between $\mathcal{ABA}84$ and $\mathcal{ABAH}844$. From left to
right: the 4 inner planets, the 4 outer planets and the whole Solar System.
The $x$-axis represents the cost ($\tau/s$) and the $y$-axis the maximum energy
variation for one integration with constant step-size $\tau$.
}\label{FigABAH8402}
\end{figure}

We have just seen that for Heliocentric coordinates we can adapt the splitting
schemes described is Section~\ref{SymIntTST} using Eqs.~\ref{SymHelio01}
and~\ref{eq:leaptric}. But with this an extra term $\eps^3\tau^2$ appears in
the error approximation that will limit the performance for high-order schemes.
One can check that this error terms is associated to the algebraic expression
$b_1^3 + b_2^3 + \dots + b_n^3$. We can add an extra stage to the scheme so
that it also satisfies: $$b_1^3 + b_2^3 + \dots + b_n^3 = 0,$$ leading to
symplectic schemes with the same generalised order as before for Heliocentric
coordinates.
Table~\ref{TAB_ABAH_high} collects the coefficients for the $\mathcal{ABAH}$
scheme of order $(8,4)$ for Heliocentric coordinates.
This scheme has the same effective order as the McLachlan $\mathcal{ABA}$
scheme of order $(8,4)$ (Section~\ref{McLahan84})
but it is specific for Heliocentric coordinates.
In Figure~\ref{FigABAH8402} we compare the performance of this new scheme
against the $\mathcal{ABA}84$ scheme. As we can see, for the outer planets the
new $\mathcal{ABAH}844$ scheme behaves better that the $\mathcal{ABA}84$ for
small step-sizes.


\subsection{$\mathcal{ABAH}$ specific methods with arbitrary order $(s1,s2, ...)$}
\label{BlanesSplitHelio}

In the same way we can add the extra constraint $b_1^3 + \cdots + b_n^3 = 0$ to
the high-order schemes in Section~\ref{BlanesSplit} to have high order
splitting schemes specific for Heliocentric coordinates.
In Table~\ref{TAB_ABAH_high} we show the coefficients of two $\mathcal{ABAH}$
schemes of orders $(8,6,4)$ and $(10,6,4)$. All these schemes have one more
stage than the schemes presented in Table~\ref{BlanesMethod}.

\begin{table}
\caption{
Coefficients for $\mathcal{ABAH}$ specific symmetric splitting methods for
Heliocentric coordinates of orders (8,4), (8, 6, 4) and (10, 6, 4) \citep{Bla12}.
}\label{TAB_ABAH_high}
\centering
\begin{tabular}{cccl}
\hline\noalign{\smallskip}
{\tt id} & {\tt order} & {\tt stages} & $\ \ a_i, b_i$ \\
\noalign{\smallskip}\hline\noalign{\smallskip}
{\tt ABAH844} & $(8,4)$ & 6 &
{\footnotesize \begin{tabular}{rcl}
$a_1$ &=& {\tt \ 0.27414026894340187616405654402  } \\ 
$a_2$ &=& {\tt  -0.10756843844016423062511052968 } \\ 
$a_3$ &=& {\tt -0.04801850259060169269119541721 } \\ 
$a_4$ &=& {\tt \  0.76289334417472809430449880574 } \\
$b_1$ &=& {\tt \ 0.64088579516251271773224911649}  \\
$b_2$ &=& {\tt  -0.85857544895678285658812832469}  \\
$b_3$ &=& {\tt \ 0.71768965379427013885587920820}
\end{tabular} } \\
\noalign{\smallskip}\hline\noalign{\smallskip}
{\tt ABAH864} & $(8,6,4)$ & 8 &
{\footnotesize \begin{tabular}{rcl}
$a_1$ &=& {\tt \ 0.06810235651658372084723976682} \\
$a_2$ &=& {\tt \ 0.25113603872210332330728295804} \\
$a_3$ &=& {\tt  -0.07507264957216562516006821767} \\
$a_4$ &=& {\tt  -0.00954471970174500781148821895} \\
$a_5$ &=& {\tt \ 0.53075794807044717763406742353} \\
$b_1$ &=& {\tt \ 0.16844325936189545343103826977} \\ 
$b_2$ &=& {\tt \ 0.42431771737426772243003516574} \\ 
$b_3$ &=& {\tt  -0.58581096946817568123090153554} \\ 
$b_4$ &=& {\tt \ 0.49304999273201250536982810002} \\ 
   \end{tabular} }\\
\noalign{\smallskip}\hline\noalign{\smallskip}
{\tt ABAH1064} & $(10,6,4)$ & 9 &
{\footnotesize \begin{tabular}{rcl}
$a_1$ &=& {\tt \ 0.04731908697653382270404371796} \\
$a_2$ &=& {\tt \ 0.26511052357487851595394800361} \\
$a_3$ &=& {\tt  -0.00997652288381124084326746816} \\
$a_4$ &=& {\tt  -0.05992919973494155126395247987} \\
$a_5$ &=& {\tt \ 0.25747611206734045344922822646} \\
$b_1$ &=& {\tt \ 0.11968846245853220353128642974}\\ 
$b_2$ &=& {\tt \ 0.37529558553793742504201285376}\\ 
$b_3$ &=& {\tt  -0.46845934183259937836508204098}\\ 
$b_4$ &=& {\tt \ 0.33513973427558970103930989429}\\ 
$b_5$ &=& {\tt \ 0.27667111912108009750494572633}\\ 
   \end{tabular} }\\
\noalign{\smallskip}\hline
\end{tabular}
\end{table}

\begin{figure}[h!]
\centering
\includegraphics[width = 0.32\textwidth]{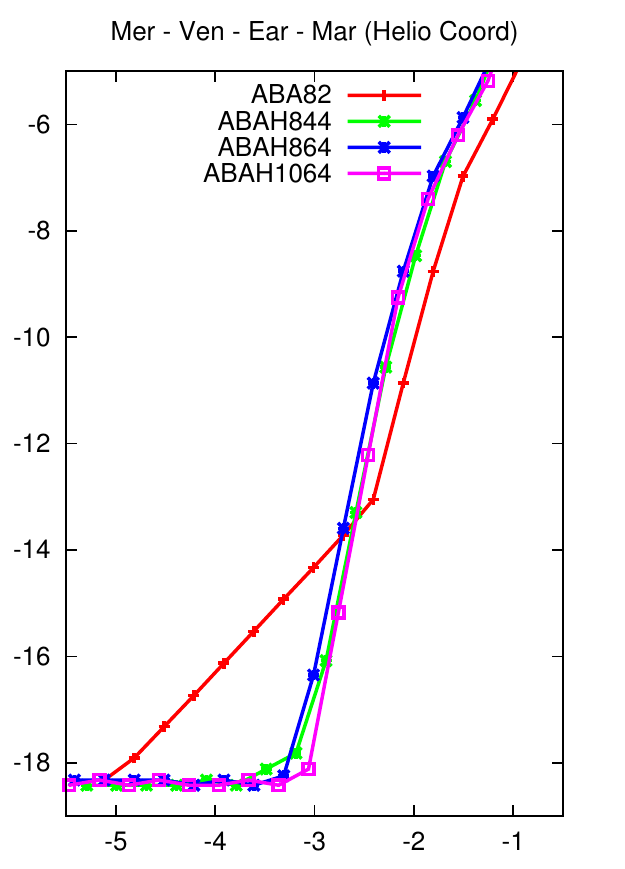}
\includegraphics[width = 0.32\textwidth]{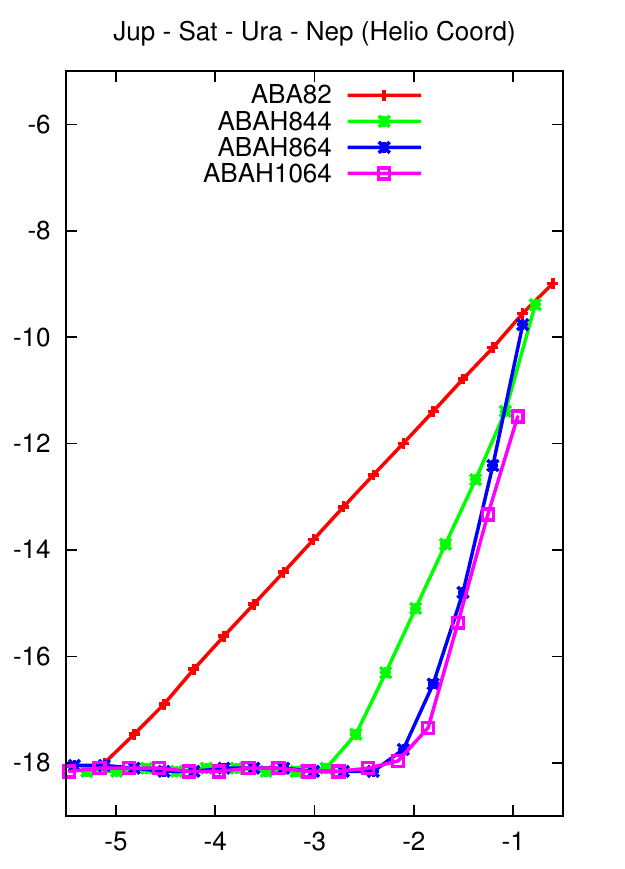}
\includegraphics[width = 0.32\textwidth]{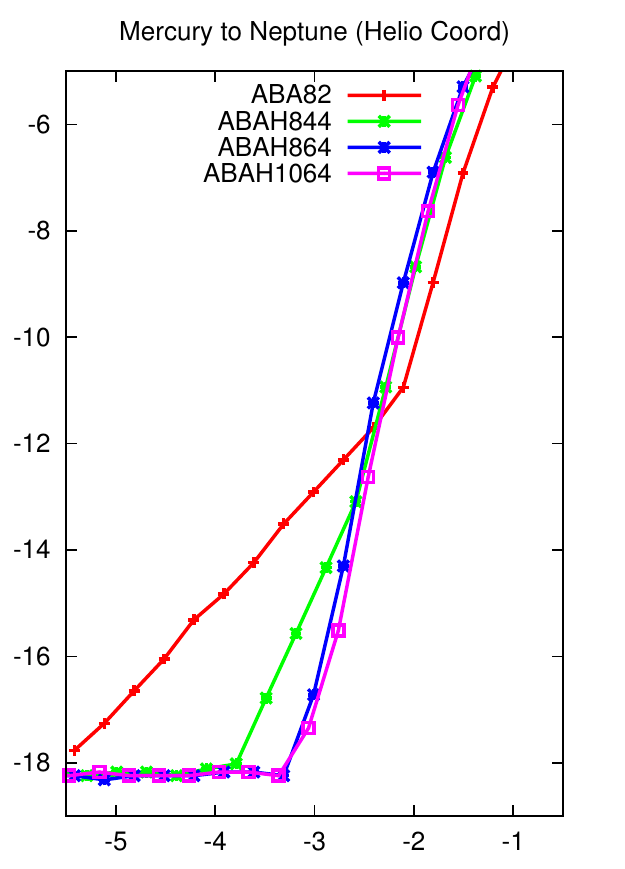}
\caption{
Comparison between $\mathcal{ABAH}$ schemes of order $(8,4,4)$, $(8,6,4)$ and
$(10,6,4)$ specific for Heliocentric coordinates and the $\mathcal{ABA}82$. From
left to right: the 4 inner planets, the 4 outer planets and the whole Solar System.
The $x$-axis represents the cost ($\tau /n$) and the $y$-axis the maximum energy
variation for one integration at constant step-size $\tau$.
}\label{Fig_ABAH_high}
\end{figure}

In Figure~\ref{Fig_ABAH_high} we compare the performance of the
$\mathcal{ABA}82$ with the other three schemes in Table~\ref{TAB_ABAH_high}. Where 
the behaviour of the schemes depending on its order is similar to the one presented
in Jacobi coordinates.
For the inner planets (Figure~\ref{Fig_ABAH_high} left) all $\mathcal{ABAH}$
schemes present a similar optimal cost. For the outer planets (Figure~\ref{Fig_ABAH_high}
middle) the $\mathcal{ABAH}$ schemes of order $(8,6,4)$ and $(10,6,4)$ are much
better than the other two schemes. We recall that here the size of the perturbation
is larger and killing the terms of order $\eps^3\tau^{4}$ does make a difference.
Finally, if we consider the 8 planets in the Solar System (Figure~\ref{Fig_ABAH_high}
right) here the $\mathcal{ABAH}$ schemes of order $(8,6,4)$ and $(10,6,4)$ do improve
the performance of the schemes of order $(8,4)$. We recall that this was
not the case in Jacobi coordinates (Figure~\ref{FigABAHigh}).

\section{Jacobi vs Heliocentric coordinates} \label{SymIntHvsJ}

In Sections~\ref{SymIntTST} and~\ref{SymIntHelio} we have described different
splitting schemes for both Jacobi and Heliocentric set of coordinates. In the
case of Heliocentric coordinates the expressions for the Hamiltonian are less
cumbersome and easier to handle (see Appendix~\ref{IntScheme}). But the size
of the perturbation is larger than in Jacobi coordinates and an extra stage
to deal with the non-integrability of $H_I$ must be added to have high-order
schemes. Here we want to compare the performance of the different schemes for 
both set of coordinates. 

\begin{figure}[h!]
\includegraphics[width = 0.32\textwidth]{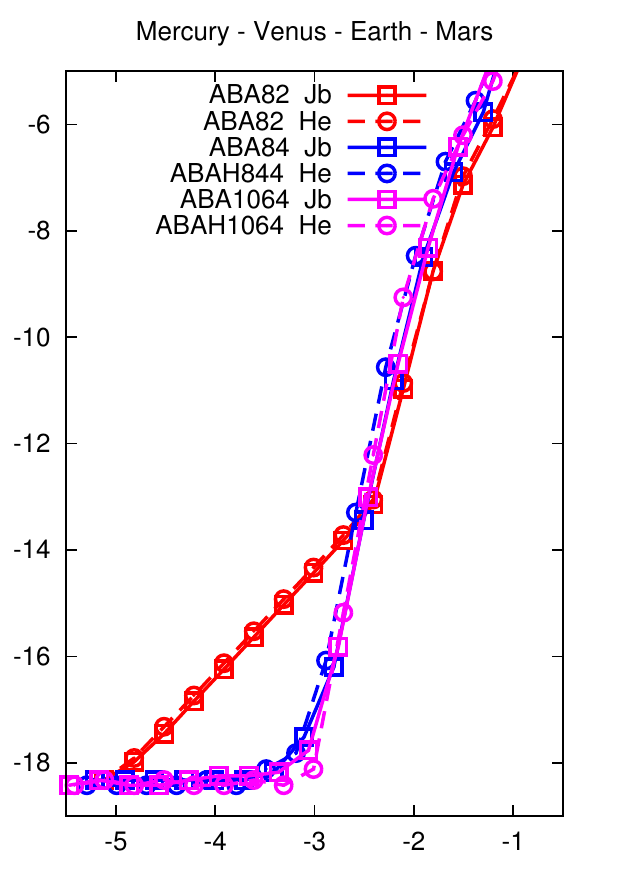}  
\includegraphics[width = 0.32\textwidth]{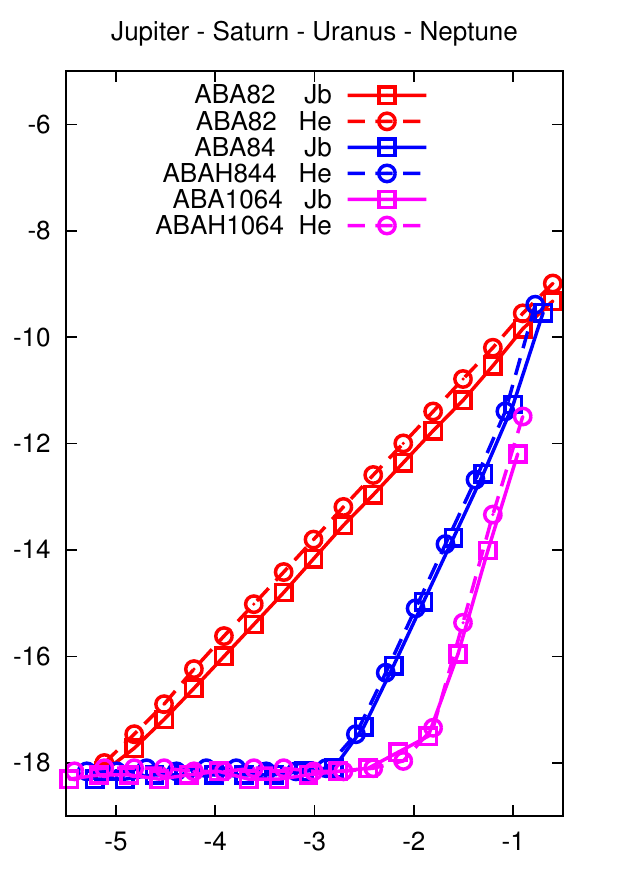}  
\includegraphics[width = 0.32\textwidth]{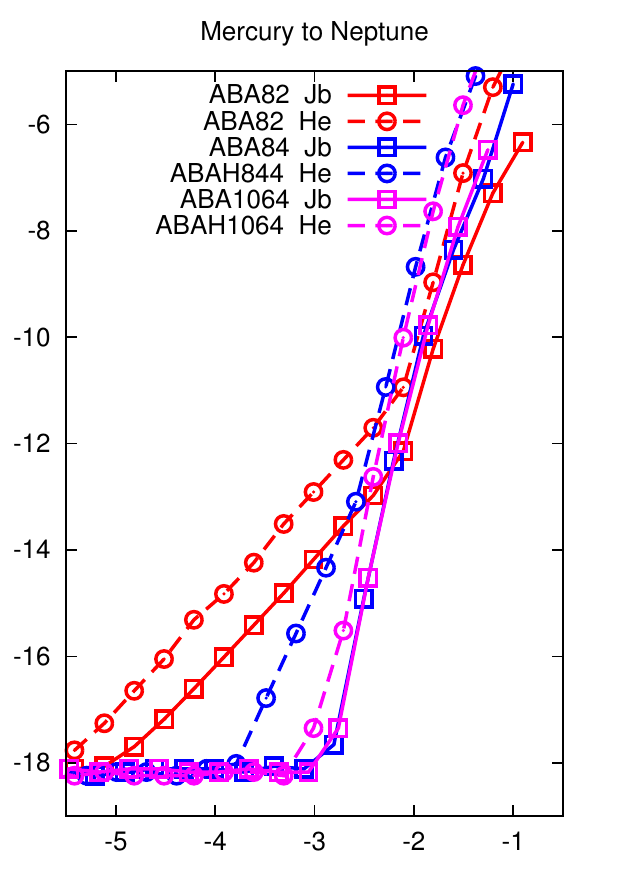}
\caption{
Comparison between Jacobi (continuous lines) and Heliocentric (discontinous lines) 
coordinates using the schemes $\mathcal{ABA}82$ (red), $\mathcal{ABA}84$ (blue) and 
$\mathcal{ABA}1064$ (purple). From left to right: the 4 inner planets, the 4 outer 
planets and the whole Solar System. The $x$-axis represents the cost ($\tau/s$) of the 
method and the $y$-axis the maximum energy variation for one integration with constant 
step-size $\tau$.
}\label{JacvsHelio}
\end{figure}

We compare the performance of the $\mathcal{ABA}$ methods of order $(8,2)$, 
$(8,4)$ and $(10,6,4)$ in both set of coordinates, with the three test models 
used throughout this report. We recall that the $(8,4)$ and $(10,6,4)$ schemes 
have an extra stage in Heliocentric coordinates. 
In Figure~\ref{JacvsHelio} we summarise the performance of these schemes. From
left to right we have the results for the inner planets, the outer planets and
the whole Solar System. We distinguish the order of the schemes by the colour.
Where the lines in red represent the schemes of order $(8,2)$, the blue lines
those of order $(8,4)$ and the purple lines those of order $(10,6,4)$. We 
use continuous lines to refer to Jacobi coordinates and discontinuous lines for 
Heliocentric coordinates. 

If we look at the results for the inner planets (Figure~\ref{JacvsHelio} left), 
we can see there is not much difference between taking Jacobi or Heliocentric 
coordinates (continuous vs discontinuous lines). In both cases the size of the 
perturbation is small (Table~\ref{TPertSize}) and there is not much difference 
between taking a splitting scheme of order $(8,4)$ or $(10,6,4)$. In both cases 
the terms in $\eps$ in the error expansion are the ones that matter, but there 
is not much difference between taking order $8$ or $10$ in $\eps\tau^k$. We 
should have to use arithmetics with higher precision to see the difference 
(see Appendix~\ref{Apndx_QP}). Hence the $\mathcal{ABA}84$ is the best
choice for this case. 

If we look at the results for the outer planets (Figure~\ref{JacvsHelio}
middle), again there is no significant difference between Jacobi and
Heliocentric coordinates. But here the $\mathcal{ABA}$ schemes of order
$(10,6,4)$ performs much better that the other schemes, having an optimal
step-sizes one order of magnitude larger than the ones for the schemes of order
$(8,4)$. 

If we look at the whole Solar System (Figure~\ref{JacvsHelio} right), we see
that there is a big difference between taking Jacobi or Heliocentric
coordinates.  Looking at the $\mathcal{ABA}82$ scheme (red lines) we see that
the slopes are the same but that there is a difference of about one order of
magnitude in accuracy for a given step-sizes. 
If we look at the scheme of order $(8,4)$ (blue lines), we see that in Jacobi
coordinates the methods behaves as one of order $8$, while in Heliocentric
coordinates this one behaves as one of order $4$. This can be explained by the
difference in the size of the perturbation (see Table~\ref{TPertSize}) in both
set of coordinates. We also see that there is a big difference between the
optimal step-size for both set of coordinates, making Jacobi coordinates by far 
the best choice. 
Finally, if we compare the $\mathcal{ABA}$ schemes of order $(10,6,4)$ (purple
lines) the difference between the two set of coordinates is drastically
reduced, although Jacobi coordinates still perform slightly better. 
While the extra stages to go from an $(8,4)$ scheme to a $(10,6,4)$ one do not 
improve in Jacobi coordinates. This is not the case of Heliocentric coordinates, 
where the $(10,6,4)$ gives the best results and the difference between the two 
set of coordinates is not as relevant as before. 

Although Jacobi coordinates presents better results for most of the test
models, we believe that using Jacobi or Heliocentric coordinates is a matter of
choice.


\section{Conclusions} 

In this article we have reviewed different symplectic splitting schemes and
tested their performance for the case of the planetary motion. We recall that 
in the case of the planetary motion, using an appropriate change of variables, 
the Hamiltonian of the N~-~body problem can be rewritten as $H_K + H_I$. A sum 
of independent Keplerian motions for each planet ($H_K$) and a small 
perturbation term given by the interaction between the planets ($H_I$). 

There are two set of canonical coordinates that allow us to write the
Hamiltonian in this way: Jacobi and Heliocentric coordinates
(Section~\ref{NBP_coord}). Although in Jacobi coordinates the size of the
perturbation is smaller and $H_I$ is integrable, Heliocentric coordinates seem
more natural and the expressions are easier to handle
(Appendix~\ref{IntScheme}). In this article we have compared the 
performance of different symplectic splitting schemes in both set of
coordinates. 

In Section~\ref{SymIntTST} we described different splitting symplectic schemes
for Jacobi coordinates. In Section~\ref{SymIntHelio} we saw how to extend these
schemes to use Heliocentric coordinates. We note that all the splitting schemes 
for Jacobi coordinates can also be used in Heliocentric coordinates, but 
in order to have a comparable performance an extra stage to kill the terms 
in $\eps^3\tau^2$ must be added (see Section~\ref{SymIntHelio}).
 
We have seen that in Jacobi coordinates, the $\mathcal{ABA}84$ scheme
introduced by \cite{McLa9502} and the $\mathcal{ABA}1064$ scheme \cite{Bla12} 
give the best results when we look at the motion of the whole Solar System. 
The high eccentricity of Mercury and its fast orbital period are the main
limiting factors and taking higher order splitting schemes do not always 
provide significant improvments. But for different planetary configurations, 
as the 4 outer planets, the $\mathcal{ABA}1064$ has a better performance than 
the $\mathcal{ABA}84$. 

When we consider Heliocentric coordinates, the
$\mathcal{ABAH}1064$ \citep{Bla12} gives the best results when 
we consider the whole Solar System. In this case, probably because the size of the
perturbation is larger, adding extra stages to have higher order schemes does
improve the results. 

Moreover,  the performances of the schemes in both set of
coordinates, Jacobi or Heliocentric are very similar for the scheme of order $(10,6,4)$, 
with a slight advantage for the Jacobi coordinates. Depending on the problem, 
one can thus use either system of coordinates, but it is clear 
that using high order schemes as the $\mathcal{ABA(H)}864$ and
$\mathcal{ABA(H)}1064$ \cite{Bla12} can drastically improve the results.  
This should be even more the case  for highly perturbed systems as 
some extra solar planetary systems with close planets of large masses.

\clearpage
\addcontentsline{toc}{section}{APPENDIX}
\begin{appendix} 

\section{On the Compensated Summation}
\label{Apndx_CS}

Using any of the symplectic integrating schemes described in this article, we 
require successive evaluations of $\exp(a_i\tau A)$ and $\exp(b_i\tau B)$. Each 
of these evaluations slightly modifies the position and velocity of each planet. 
For $\tau$ small, we will have a loss in accuracy due to round-off errors. The 
{\it Compensated Summation} is a simple trick that is commonly used to reduce 
the round-off error.
In a general framework, when we consider a numerical method for solving an ODE, 
we require a recursive evaluation of the form:
\begin{equation}
y_{n+1} = y_n + \delta_n,
\label{EqCS01}
\end{equation}
where $y_n$ is the approximated solution and $\delta_n$ is the increment to be done. 
Usually $\delta_n$ will be smaller in magnitude than $y_n$. In this situation, the 
rounding errors caused by the computation of $\delta_n$ are in general smaller that
those to evaluate Eq. $\ref{EqCS01}$. The algorithm that can be used in order to 
reduce this round-off error is called the {\it ``Compensated Summation''} 
\citep{Kahan}. 

\vspace{3mm}
{\bf Compensated Summation Algorithm:}  
{\it Let $y_0$ and $\{\delta_n\}_{n\geq 0}$ be given and put $e=0$. Compute 
$y_1,y_2, \dots$ from Eq. \ref{EqCS01} as follows: }\\
$$\begin{array}{l}
for\  n = 0, 1, 2, \dots \ do \\ 
	\qquad a = y_n \\
	\qquad e = e + \delta_n \\
	\qquad y_{n+1} = a + e \\
	\qquad e = e + (a - y_{n+1})\\
end do
\end{array}
$$
\vspace{3mm}

\begin{figure}[!h]
\centering
\includegraphics[width = 0.48\textwidth]{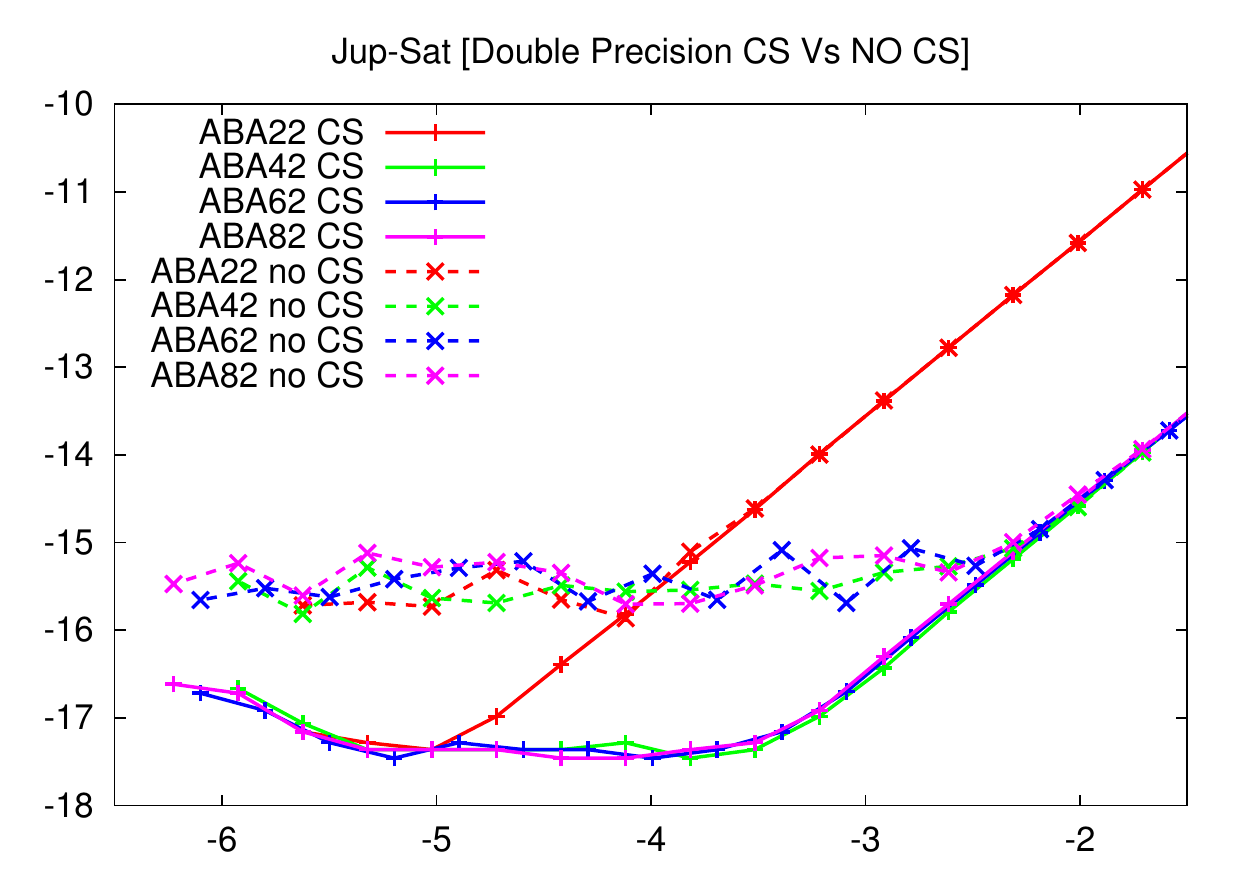} 
\includegraphics[width = 0.48\textwidth]{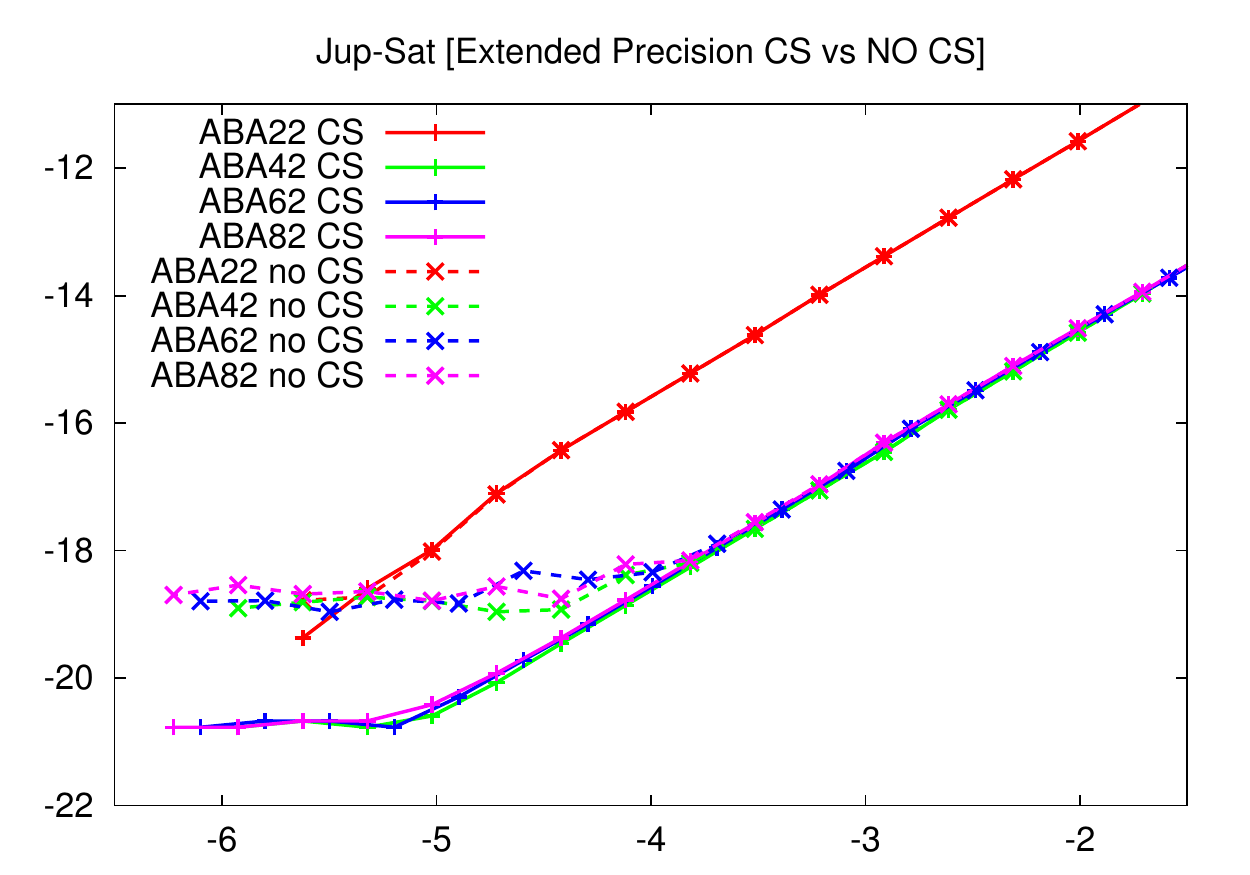}
\caption{Comparison between the $\mathcal{ABA}$ schemes of order $(2n,2)$ for
$n=1,2,3,4$ applied to the Sun-Jupiter-Saturn three body problem. With (CS) and
without (noCS) the compensated summation. The $x$-axis represent the cost
$(\tau/n)$ and the $y$-axis the maximum energy variation for one integration
with constant step-size $\tau$. Left: using a double precision arithmetics;
Right: using an extended precision arithmetics.}\label{FigSC01}
\end{figure}

This algorithm accumulates the rounding errors in $e$ and feeds them back into
the summation when possible. At each time-step of the integration, when we
evaluate $\exp(a_i\tau A)$ or $\exp(b_i\tau B)$, the increment in position and
velocity is done using the compensated summation.  
In Figure \ref{FigSC01} we show the results for the $\mathcal{ABA}$ $(2n,2)$
schemes for $n = 1,2,3,4$ using double (left) and extended precision (right).
In both cases we gain almost one order of magnitude in precision when we take
into account the compensated summation. 

\section{Integration Schemes ({\it some help on the practical coding})}
\label{IntScheme}

In this paper we have reviewed many splitting symplectic integrating schemes, 
all of them of the form: 
\begin{equation}
\mathcal{S}(\tau) = \exp(a_1\tau A) \exp(b_1\tau B) \ \dots \ 
	\exp(b_1\tau B) \exp(a_1\tau A), 
\label{SABAscheme01}
\end{equation}
where $\exp(\tau A)$ and $\exp(\tau B)$ can be computed explicitly. They 
correspond to the integrals of the two different parts of the original
Hamiltonian. 
In this section we show how to compute explicitly $\exp(\tau A)$ and
$\exp(\tau B)$ for the particular case of the N-body problem in Jacobi and
Heliocentric coordinates. We note that from now on: $\bf \tilde{u}$ stands for 
the momenta associated to ${\bf u}$ and $\bf u'$ stands for $d{\bf u}/dt$. 

\subsection{Keplerian Motion ($H_{K}$)}

We recall that in {\bf Jacobi coordinates}, 
\begin{equation}
H_{K} = \sum_{i=1}^{n} \left( 
  \frac{1}{2} \frac{\eta_i}{\eta_{i-1}}\frac{||{\bf \tilde v_i}||^2}{m_i} 
  - G\frac{m_i \eta_{i-1}}{||{\bf v_i}||} 
  \right),
\label{KepJacob}
\end{equation}
whereas in {\bf Heliocentric coordinates},  
\begin{equation}
H_{K} = \sum_{i=1}^{n} \left( 
  \frac{1}{2} ||{\bf \tilde r_i}||^2 \left[\frac{m_0 + m_i}{m_0  m_i}\right] 
  - G\frac{m_0 m_i}{||{\bf r_i}||} 
      \right). 
\label{KepHelio}
\end{equation}
In both cases $H_K$ is a sum of independent Keplerian motions. In Jacobi 
coordinates each planet follows an elliptical orbit around the centre on mass 
of the Sun and the planets that are closer to the Sun, the mass parameter of 
the system is $\mu_J = G\eta_i$. While in Heliocentric coordinates each planet 
follows an elliptical orbit around the planet-Sun centre of mass, and the mass 
parameter of the system is $\mu_H = G(m_0+m_i)$. 

It is well known that Kepler problem is integrable, but the solution from time 
$t = t_0$ to $t = t_0 + \tau$ is expressed in a simple form if we consider 
action-angle variables. To compute $\exp(\tau L_{H_K})$ we need to be able to 
compute $\bf(r(t_0+\tau),v(t_0+\tau))$ from $\bf(r(t_0), v(t_0))$. 

An option is to change to elliptical coordinates, advance the mean anomaly and
then return to cartesian coordinates. But this can accumulate a lot of
numerical errors as well as it is very expensive in terms of computational
cost. Instead we use a similar idea as the Gauss $f$ and $g$
functions~\citep{Danby92}, where we use an expression for the increment in
position and velocities for a given step-size $\tau$, without having to perform
any change of coordinates. Let us give some details on how to derive these
expressions. 

In elliptical coordinates the motion of the two body problem is given by
$(a,e,i, \Omega, \omega, E)$, where all of the elements remain fixed except for
$E$ that varies following Kepler equation ($n(t -t_p) = M = E - e\sin E$).
Using a reference frame where the orbital plane is given by $Z=0$, the $X$-axis
is the direction of the perihelion and the $Y$-axis completes an orthogonal
reference system on the orbital plane, the position $(X, Y, 0)$ and velocity
$(X', Y', 0)$ are given by:
\begin{equation}
\begin{array}{rclcrcl}
	X &=& a(\cos E - e), & \phantom{qqq} &  Y &=& a\sqrt{1-e^2} \sin E,	\\ 
	\\
	X' &=& -\displaystyle\frac{na^2}{r}\sin E,  & \phantom{qqq} & 
		Y' &=& \displaystyle\frac{na^2}{r}\sqrt{1-e^2} \cos E,
\end{array}
\end{equation}
where $r = a(1 - e \cos E)$ and $n = \mu^{1/2}a^{-3/2}$. 
The position and velocities on a fixed reference frame are given by: 
\begin{equation}
\left(\begin{array}{cc}
x & x' \\ y & y'\\ z  & z'
\end{array}\right) = 
\mathcal{R}_3(\Omega)\times \mathcal{R}_1(i)\times \mathcal{R}_3(\omega)
\times 
\left(\begin{array}{cc}
X & X' \\ Y & Y'\\ 0  & 0
\end{array}\right),  \label{eqchange1}
\end{equation}
where 
$$
\mathcal{R}_1(\theta) = 
\left(\begin{array}{ccc}
1 \ & \ 0 \ & \ 0 \\ 
0 \ & \ \cos\theta \  & \ \sin\theta \\ 
0 \ & \ -\sin\theta\  & \ \cos\theta 
\end{array}\right), 
\mbox{  and   }
\mathcal{R}_3(\theta) = 
\left(\begin{array}{ccc}
  \cos\theta \  & \ \sin\theta \  & \  0 \\ 
 -\sin\theta \  & \ \cos\theta \  & \  0 \\
 0 \ & \ 0 \ & \ 1 \\ 
\end{array}\right)
$$
Notice that  
$$\mathcal{R}_3(\Omega)\times \mathcal{R}_1(i)\times \mathcal{R}_3(\omega) 
= \mathcal{R} \times \mathcal{R}_3(\varpi),$$
where $\varpi = \Omega + \omega$ and $\mathcal{R} = \mathcal{R}_3(\Omega)\times 
\mathcal{R}_1(i)\times \mathcal{R}_3(-\Omega)$. Given that $\mathcal{R}_1(i) = 
\mathcal{R}_1(i/2)\mathcal{R}_1(i/2)$ we have that: 
\begin{equation}
\mathcal{R} = 
\left(\begin{array}{ccc}
1 - 2{\bf p}^2  & 2{\bf p q}       &  2{\bf p} \chi \\
2{\bf p q}      & 1 - 2{\bf q}^2   & -2{\bf q} \chi \\
-2{\bf p} \chi  & 2{\bf q} \chi    & 1 - 2{\bf p}^2 - 2{\bf q}^2
\end{array}\right),
\end{equation}
where ${\bf p} = \sin i/2 \sin \Omega, {\bf q} =  \sin i/2 \sin \Omega,$
and $\chi = \sqrt{1- {\bf p}^2 - {\bf q}^2} = \cos i/2$. From Eq.~\ref{eqchange1} we have, 
\begin{eqnarray}
{\bf [\ r(t_0),\ v(t_0)\ ]} &=& 
\mathcal{R}\times \mathcal{R}_3(\varpi) 
\times 
\left[
\begin{array}{cc}
X_0 & X_0' \\ Y_0 & Y_0' \\ 0 & 0
\end{array}
\right] \\
{\bf [\ r(t_0+\delta t),\ v(t_0 + \delta t)\ ]} &=& 
\mathcal{R}\times \mathcal{R}_3(\varpi) 
\times 
\left[
\begin{array}{cc}
X_1 & X_1' \\ Y_1 & Y_1' \\ 0 & 0
\end{array}
\right].
\end{eqnarray}
Hence,  
\begin{eqnarray}
{\bf [\ r(t_0 + \delta t),\ v(t_0+\delta t)\ ]} &=& 
{\bf [\ r(t_0), v(t_0)\ ]} 
\left[ 
\begin{array}{cc}
X_0 & X_0' \\ Y_0 & Y_0' 
\end{array}
\right]^{-1}
\left[
\begin{array}{cc}
X_1 & X_1' \\ Y_1 & Y_1' 
\end{array}
\right]  \\ 
&=& 
{\bf [\ r(t_0), v(t_0)\ ]} \left[
\begin{array}{cc}
a_{11} & a_{12} \\
a_{21} & a_{22}
\end{array}
\right].
\label{trans}
\end{eqnarray}
One can check that,
\begin{equation}
\begin{array}{rcl}
a_{11} &=&\displaystyle  1 + (\cos(E_1 - E_0) - 1)\frac{a}{r_0}, \\ \\
a_{21} &=&\displaystyle  \frac{a^{3/2}}{\mu^{1/2}}\sin(E_1 - E_0) - e\sin E_1 + e\sin E_0 ,\\ \\
a_{12} &=&\displaystyle  -\frac{\sqrt{a}}{r_0r_1}\sin(E_1 - E_0), \\ \\
a_{22} &=&\displaystyle  1 + (\cos(E_1 - E_0) - 1)\frac{a}{r_1}, 
\end{array} \label{aij00}
\end{equation}
where $r_{i} = a(1 - e\cos E_{i})$ for $i=0,1$. We use Kepler's equation to compute 
$\delta E = E_1-E_0$ from $\delta t = t_1 - t_0$. 
Taking $M_{i} = n(t_{i} - t_p)$ for $i=0,1$, we have that $\delta E$ is the solution of 
\begin{equation}
x - e \cos E \sin x - e \sin E \cos x + e \sin E - n\delta t = 0. \label{kep1}
\end{equation}

Calling $C = \cos \delta E,\ S = \sin \delta E$ and $ce = e \cos E_0,\ se = \sin E_0$ 
we have that $r_1 = a(1 - ce\cdot C + se\cdot S)$. Now we can rewrite Eq.~\ref{aij00} 
as: 
\begin{equation}
\begin{array}{rcl}
a_{11} &=&\displaystyle  1 + ( C - 1)\frac{a}{r_0},\\ \\
a_{21} &=&\displaystyle  \delta t + ( S - \delta E )\frac{a^{3/2}}{\mu^{1/2}}, \\ \\
a_{12} &=&\displaystyle  - \frac{S}{r_0\sqrt{a}(1 - ce\cdot C + se\cdot S)}, \\ \\
a_{22} &=&\displaystyle  1 + \frac{C - 1}{1 - ce\cdot C + se\cdot S}. 
\end{array} \label{aij11}
\end{equation}

To summarise, given ${\bf r = r(t_0), v = v(t_0)}$ and defining $r_0 = ||{\bf r}||$ 
and $v_0 = ||{\bf v}||$. We find 
$$
a = r_0/(2 - r_0v_0^2), \qquad 
ce = e \cos E_0 = r_0 v_0^2 - 1, \qquad 
se = e \sin E_0 = \langle {\bf r}, {\bf v}\rangle /\sqrt{\mu a}.   
$$
Then we take $\delta t$ and we use Eq.~\ref{kep1} to find $\delta E$. Finally we use 
Eqs.~\ref{trans} and~\ref{aij11} to find ${\bf r(t_0 + \delta t), v(t_0 + \delta t)}$.

\subsection{Jacobi Coordinates}

We recall that in this set of coordinates the perturbation part is given 
by:
\begin{equation}
H_{I} = U_1 =
  G \left[\sum_{i=1}^{n}m_i\left(\frac{\eta_{i-1}}{||{\bf v_i}||} 
 - \frac{m_0}{||{\bf r_i}||}\right) 
 - \sum_{0 < i < j\leq n}\left( \frac{m_i m_j}{||{\bf r_i - r_j ||}} \right) \right].
\end{equation}

\subsubsection{Computing $\exp(L_{H_{I}})$:}

$U_1$ depends only on the position, hence the equations of motion are given by, 
$$
\frac{d}{dt}{\bf v_k} = \frac{\partial U_1}{\partial \ \bf \tilde v_k}, 
\qquad \qquad
\frac{d}{dt}{\bf \tilde v_k} = -\frac{\partial U_1}{\partial \ \bf v_k}.  
$$
Using ${\bf \tilde v_i} =
\displaystyle\frac{\eta_{i-1}m_i}{\eta_i}{\bf \dot v_i}$ we have 
$$
{\bf v_k}(\tau) = {\bf  v_k}(\tau_0), \qquad \qquad 
{\bf \dot v_k}(\tau) = {\bf \dot v_k}(\tau_0) - 
	\tau \frac{\eta_i}{\eta_{i-1}m_i}\frac{\partial U_1}{\partial {\bf v_k}}.
$$

As the expressions for ${\partial U_1}/{\partial {\bf v_k}}$ can be a little 
cumbersome, we compute them separately. When we derive $H_{I}$ with respect to  
${\bf v_k}$ we must derive 3 main expressions: 
${1}/{||{\bf v_i}||}$, ${1}/{||{\bf r_i}||}$ and ${1}/{||{\bf r_i - r_j}||}$
for $i<j$. We first give the derivatives of these factors 
with respect to ${\bf v_k}$ and then we will deduce  
${\partial H_{I}}/{\partial \bf v_k}$ for $k = 1, \dots, n$. 

$$
\begin{array}{lcl}
\displaystyle\frac{\partial}{\partial {\bf v_k}}
    \left(\frac{1}{||{\bf v_i}||}\right) = 
  -\frac{{\bf v_i}}{||{\bf v_i}||^3}\cdot \delta_{i,k}\ , 
&\mbox{where}&
 \delta_{i,k} = 
\left\{\begin{array}{ccl} 0 &\mbox{ if } & i \neq k, \\
  1 &\mbox{ if }& i = k.\end{array}\right.
\\ \\
\displaystyle\frac{\partial}{\partial {\bf v_k}}
    \left(\frac{1}{||{\bf r_i}||}\right) = 
  -\frac{{\bf r_i}}{||{\bf r_i}||^3}\cdot \xi_{i,k}, 
&\mbox{  where  }&
\xi_{i,k} = 
\left\{\begin{array}{ccl} 0 &\mbox{ if } & i < k, \\
  1 &\mbox{ if }& i = k, \\ 
  \displaystyle\frac{m_k}{\eta_k} &\mbox{ if }& i > k.\end{array}\right.
\\ \\
\displaystyle\frac{\partial}{\partial {\bf v_k}}
    \left(\frac{1}{||{\bf r_i - r_j}||}\right) = 
  -\frac{\bf r_i - r_j}{||{\bf r_i - r_j}||^3}\cdot \psi_{i,j,k}, 
&\mbox{  where  }& \psi_{i,j,k} = 
\left\{\begin{array}{ccl} 
  \displaystyle\frac{\eta_{k-1}}{\eta_k} &\mbox{ if }& k = i < j, \\ 
  -\displaystyle\frac{m_{k}}{\eta_k} &\mbox{ if }& i < k < j, \\ 
  -1 & \mbox{ if }& i < j = k, \\  
	0 & \mbox{ else } & (k < i < j, \ \ i < j < k).
\end{array}\right.
\end{array}
$$

To compute ${\partial U_1}/{\partial \bf v_k}$ for $k = 1, \dots, n$, 
we consider separately the cases $k = 1$ and $k > 1$: 

\begin{eqnarray*} 
\frac{\partial U_1}{\partial \bf v_1} &=& 
G \frac{m_0m_1}{\eta_1}\left[
   \sum_{i=2}^{n} m_i \frac{\bf r_i}{||{\bf r_i}||^3} 
 + \sum_{i=2}^n m_i \frac{\bf r_1-r_i}{||{\bf r_1 - r_i}||^3} 
\right].
\\ && \\
\frac{\partial U_1}{\partial \bf v_k} &=& 
 Gm_k\left[
  -\eta_{k-1}\frac{\bf v_k}{||{\bf v_k}||^3}
  + m_0\frac{\bf r_k}{||{\bf r_k}||^3} 
  + \frac{m_0}{\eta_k}\sum_{i=k+1}^{n}m_i \frac{\bf r_i}{||{\bf r_i}||^3}
\right. \\ 
&+& \left.
  \frac{\eta_{k-1}}{\eta_k}
\sum_{j=k+1}^n  m_j\frac{\bf r_k - r_j}{||{\bf r_k - r_j}||^3} 
- \sum_{i=1}^{k-1} m_i \frac{\bf r_i - r_k}{||{\bf r_i - r_k}||^3} 
-\frac{1}{\eta_k}
 \sum_{i=1}^{k-1}\sum_{j=k+1}^n m_im_j \frac{\bf r_i-r_j}{||{\bf r_i - r_j}||^3}
 \right]. \\ 
\end{eqnarray*}

\subsubsection{Computing the Corrector: $\exp(L_{\{\{A,B\},B\}})$}

In Section~\ref{SABACn} we described a splitting symplectic schemed where a 
corrector term was added at the beginning and at the end of each step-size. 
The corrector term is given by, 
$$\exp(-\tau^3\varepsilon^2\frac{c}{2}L_C),$$ 
with $L_C = L_{\{\{A,B\},B\}}$ and $c$ a constant coefficient that depends 
on the order of the $\mathcal{ABA}$ scheme. 

In Jacobi coordinates $A$ is quadratic in $p$ and $B$ only depends on $q$ so
$\{\{A,B\},B\}$ only depends on $q$ and $\{\{A,B\},B\}$ is integrable. We
recall that $A = H_{Kep} = T_0 + U_0$ and $B = H_{pert} = U_1$. Hence,
$$\{\{T_0 + U_0, U_1\}, U_1\} = \{\{T_0, U_1\}, U_1\}.$$ 
Given that  
$T_0 = \displaystyle\sum_{i=1}^{n} \frac{\eta_i}{\eta_{i-1}m_i}
	\frac{||{\bf\tilde v_i} ||^2}{2}$, we have,   
\begin{eqnarray*}
\{T_0, U_1\} &=& 
	\sum_{i=1}^{n} \frac{\eta_i}{\eta_{i-1}m_i} {\bf\tilde v_i} 
		\frac{\partial U_1}{\partial \bf v_i}, \\
\{\{T_0, U_1\}, U_1\} &=& \sum_{i=1}^{n} 
   \frac{\eta_i}{\eta_{i-1}m_i} \left(\frac{\partial U_1}{\partial \bf v_i} \right)^2.  
\end{eqnarray*}
Then the equations of motion for $L_C$ are given by:
$$
\begin{array}{l}
	{\bf v_k} (\tau) = {\bf v_k}(\tau_0), \\
\displaystyle	{\bf \tilde v_k} (\tau) 
		= {\bf \tilde v_k}(t_0) + \tau \sum_{i=1}^{n} 2 \gamma_i 
			\frac{\partial U_1}{\partial \bf v_i} 
			\frac{\partial^2 U_1}{\partial \bf v_i \partial\bf v_k }, 
\end{array}
$$
where $\gamma_k = \frac{\eta_k}{\eta_{k-1}m_k}$. 
As before, using  
${\bf \tilde v_i} = \displaystyle\frac{\eta_{i-1}m_i}{\eta_i}{\bf \dot v_i}$ 
we have  
$$
{\bf \dot v_k} (\tau) = {\bf \dot v_k}(t_0) + 
	\tau \gamma_k \sum_{i=1}^{n} 2 
		\left( \gamma_i \frac{\partial U_1}{\partial \bf v_i} \right) 
		\frac{\partial^2 U_1}{\partial \bf v_i \partial\bf v_k }.
$$
Again the expression for 
$\displaystyle\frac{\partial^2 U_1}{\partial \bf v_i \partial\bf v_k}$ are a
little cumbersome and we first show how to derive the different parts in
$\displaystyle\frac{\partial U_1}{\partial \bf v_k}$:  
${\bf v_i}/{||{\bf v_i}||^3}$, ${\bf r_i}/{||{\bf r_i}||^3}$ and 
${\bf r_i - r_j}/{||{\bf r_i - r_j}||^3}$.

$$
\begin{array}{lcl}
\displaystyle  \frac{\partial}{\partial {\bf v_k}}
    \left(\frac{\bf v_i}{||{\bf v_i}||^3}\right) 
	 &= &
\displaystyle    \left(\frac{\langle{\bf h, k}\rangle}{||{\bf v_i}||^3} 
   -3\frac{\langle{\bf v_i, h}\rangle \langle{\bf v_i, k}\rangle}{||{\bf v_i}||^5}\right)
	 \cdot \delta_{i,k}, 
\\ \\ 
\displaystyle  \frac{\partial}{\partial {\bf v_k}}
    \left(\frac{\bf r_i}{||{\bf r_i}||^3}\right) 
	 &=& 
\displaystyle  \left(\frac{\langle{\bf h, k}\rangle}{||{\bf r_i}||^3}
  -3\frac{\langle{\bf r_i, h}\rangle\langle{\bf r_i, k}\rangle}{||{\bf r_i}||^5}
  \right)\cdot \xi_{i,k}, 
\\ \\
\displaystyle \frac{\partial}{\partial {\bf v_k}}
    \left(\frac{\bf r_i - r_j}{||{\bf r_i - r_j}||^3}\right) 
	 &=& 
\displaystyle  \left(\frac{\langle{\bf h, k}\rangle}{||{\bf r_i - r_j}||^3} \right. 
   \displaystyle\left. -3\frac{\langle{\bf r_i - r_j, h}\rangle
 	\langle{\bf r_i - r_j, k}\rangle}
	{||{\bf r_i - r_j}||^5} \right)\cdot \psi_{i,j,k}. 
\end{array}
$$

From now on we call 
${\bf Acc}(i) = \gamma_i \displaystyle\frac{\partial U_1}{\partial \bf v_i}$, 
and
$$
\begin{array}{rcl}
\Lambda_s &=& {\bf Acc}(s) 
  \displaystyle\left(\frac{\langle{\bf h, k}\rangle}{||{\bf v_i}||^3} 
   -3\frac{\langle{\bf v_i, h}\rangle 
		\langle{\bf v_i, k}\rangle}{||{\bf v_i}||^5}\right),	
  \\ && \\
\Theta_{i,s} &=&  {\bf Acc}(s) 
  \displaystyle\left(\frac{\langle{\bf h, k}\rangle}{||{\bf r_i}||^3}
  -3\frac{\langle{\bf r_i, h}\rangle\langle{\bf r_i, k}\rangle}{||{\bf r_i}||^5}
  \right),
  \\ && \\
\Psi_{i,j,s} &=& {\bf Acc}(s) 
  \displaystyle\left(\frac{\langle{\bf h, k}\rangle}{||{\bf r_i - r_j}||^3}
  -3\frac{\langle{\bf r_i - r_j, h}\rangle\langle{\bf r_i - r_j, k}\rangle}
  {||{\bf r_i - r_j}||^5} \right).\\
\end{array}
$$

We can now give the expressions for 
$\displaystyle\frac{\partial^2 U_1}{\partial \bf v_i \partial\bf v_k}$ $\forall j, k$
\begin{eqnarray*}
\frac{\partial^2 U_1}{\partial {\bf v_1} \partial {\bf v_1}} 
	&=& G\frac{m_0m_1}{\eta_1}
   \left[ \sum_{j = 2}^n m_j( \Theta_{j,1}\frac{m_1}{\eta_1} + 
	  \Psi_{1,j,1}\frac{m_0}{\eta_1})
	\right]. \\ 
&& \\
\frac{\partial^2 U_1}{\partial {\bf v_1} \partial {\bf v_k}}
    &=& G\frac{m_0m_1m_k}{\eta_1} \left[
    \Theta_{k,s} - \Psi_{1,k,s} 
	 + \frac{1}{\eta_k} \sum_{j = k+1}^n m_j( \Theta_{j,s} - \Psi_{1,j,s})
	\right]. \\ 
&& \\
\frac{\partial^2 U_1}{\partial {\bf v_k} \partial {\bf v_k}}
   &=& G m_k \left[
   -\eta_{k-1} \Lambda_k + m_0 \Theta_{k,k} + 
	\frac{m_0m_k}{\eta_k^2}\sum_{i = k+1}^{n} m_i \Theta_{i,k} 
	+ \frac{\eta_{k-1}^2}{\eta_k^2} \sum_{i = k+1}^{n} m_i \Psi_{k,i,k} \right. \\
	&& \left. + \sum_{i = 1}^{k-1} m_i \Psi_{i,k,k}
	+ \frac{m_k}{\eta_k^2} \sum_{i = 1}^{k-1}\sum_{j = i+1}^n m_im_j \Psi_{i,j,k}
  \right]. \\ 
&& \\
\frac{\partial^2 U_1}{\partial {\bf v_k} \partial {\bf v_l}} 
	&=& G\frac{m_k m_l}{\eta_k}\left[
   m_0 \Theta_{l,s}- \eta_{k-1}\Psi_{k,l,s}
	+ \frac{1}{\eta_l}\sum_{i = l+1}^{n} m_i (m_0\Theta_{i,s} - \eta_{k-1}\Psi_{k,i,s}) \right. \\
	&& \left. + \sum_{i = 1}^{k-1} m_i \Psi_{i,l,s} 
	+ \frac{1}{\eta_l} \sum_{i = 1}^{k-1}\sum_{j = l+1}^n m_im_j \Psi_{i,j,k}
	\right]. \\
\end{eqnarray*}

\subsection{Heliocentric Coordinates} 

We recall that in this set of coordinates the perturbation part is given by:
\begin{equation}
H_{I} = T_{1} + U_{1} = 
  \sum_{0 < i < j\leq n} \frac{\bf \tilde r_i \cdot \tilde r_j}{m_0} 
  \ - \
  G\sum_{0 < i < j\leq n} \frac{m_i m_j}{\Delta_{ij}},
\end{equation}

\subsubsection{Computing $\exp(\tau L_{T_{1}})$:}

Notice that $T_{1}$ depends only on the momenta (${\bf \tilde r}$). Hence, 
the equations of motion are given by, 
$$
\begin{array}{l}
\displaystyle \frac{d}{dt}{\bf r_k} 
	= \frac{\partial T_{1}}{\partial \ \bf \tilde r_k} 
	= \sum_{j = 1, j \neq k} \frac{\bf \tilde r_j}{m_0} 
	= \sum_{j = 1, j \neq k} \frac{m_j \bf \dot r_j}{m_0}, \\ \\
\displaystyle \frac{d}{dt}{\bf \tilde r_k} 
	= \frac{\partial T_{1}}{\partial \ \bf r_k} = 0.
\end{array}
$$
Finally, 
$$
{\bf r_k}(\tau) = {\bf r_k}(\tau_0) + 
	\tau \sum_{j = 1, j \neq k} \frac{m_j \bf \dot r_j}{m_0}, 
	\qquad \qquad 
{\bf \dot r_k}(\tau) = {\bf \dot r_k}(\tau_0).
$$

\subsubsection{Computing $\exp(\tau L_{U_{pert}})$:}

Notice that $U_{1}$ depends only on the positions (${\bf r}$). Hence, 
the equations of motion are given by, 
$$
\begin{array}{l}
	\displaystyle \frac{d}{dt}{\bf \tilde r_k} = 
		\frac{\partial U_{1}}{\partial \ \bf r_k} = 0, \\ \\
	\displaystyle \frac{d}{dt}{\bf \tilde r_k} = 
		-\frac{\partial U_{1}}{\partial \ \bf r_k} 
		= -G \left( \sum_{j = 1}^{k-1} \frac{m_k m_j}{\Delta_{kj}^3}({\bf r_k - r_j})
		- \sum_{j = k+1}^{n} \frac{m_k m_j}{\Delta_{jk}^3}({\bf r_j - r_k}) \right).
\end{array}
$$
Given that ${\bf\tilde r_k} = m_k {\bf \dot r_k}$, we have: 
$$
{\bf r_k}(\tau) = {\bf r_k}(\tau_0), \qquad 
{\bf \dot r_k}(\tau) = {\bf \dot  r_k}(\tau_0)  -  
	\tau \ G \left(\sum_{j = 1}^{k-1} \frac{m_j}{\Delta_{kj}^3}({\bf r_k - r_j})
	+ \sum_{j = k+1}^{n} \frac{m_j}{\Delta_{kj}^3}({\bf r_k - r_j}) \right).
$$

\section{Heliocentric Coordinates {\it (Alternatives for the set of equations)}}
\label{Apndx_HvsH}

The canonical Heliocentric coordinates used in
Section~\ref{SymIntHelio} are canonical and the position of each body is 
taken with respect to the position of the Sun. The position and their 
associated momenta are given by: 
$$
\left.\begin{array}{rcl}
\bf r_0 &=& \bf u_0  \\
\bf r_i &=& \bf u_i - u_0
\end{array}\right\}, 
\qquad \qquad
\left.\begin{array}{rcl}
\bf \tilde r_0 &=& \bf \tilde u_0 + \dots + \tilde u_n  \\
\bf \tilde r_i &=& \bf \tilde u_i
\end{array}\right\}.
$$
The main difference between Jacobi and Heliocentric coordinates is that in the
second set of coordinates the kinetic energy is not diagonal in the momenta.
Instead we have: 
\begin{equation}
T 	= \displaystyle\frac{1}{2}\sum_{i=0}^{n}\frac{|| {\bf \tilde u_i} ||^2}{m_i}
   = \frac{1}{2} \sum_{i=1}^{n}\frac{|| {\bf \tilde r_i} ||^2}{m_i} + 
 	\frac{1}{2} \frac{|| \sum_{i=1}^n {\bf \tilde r_i} ||^2}{m_0},
\label{TH01}
\end{equation}
which can be rewritten as:
\begin{equation}
T 	= \displaystyle\frac{1}{2}\sum_{i=0}^{n}\frac{|| {\bf \tilde u_i} ||^2}{m_i}
	= \frac{1}{2} \sum_{i=1}^{n}{|| {\bf \tilde r_i} ||^2} 
	\left[\frac{1}{m_0} + \frac{1}{m_i}\right] + 
	\sum_{0<i<j} \frac{{\bf \tilde r_i \cdot \tilde r_j}}{m_0}. \label{TH02}
\end{equation}

The extra term due to the momenta of the Sun is added to the perturbation part
and makes it depend on both position and velocities. In Section~\ref{NBP_coord}
we used Eq.\ref{TH02} to derive the Hamiltonian expression.
\cite{DuLeLee98,Cham99,Wisd06} used Eq.~\ref{TH01} instead. Here we 
discuss the main differences between the two sets of equations and compare 
the performance of the integrators presented
in this paper for both expressions.

\subsection{Two different expressions for Heliocentric coordinates} 

As we know in Heliocentric coordinates the Hamiltonian for an n-planetary 
system takes the form: 
$$H = H_{K} + T_1 + U_1,$$
a sum of Keplerian parts, a quadratic term in the momenta and the gravitational 
interaction between the other planets. Using Eq.~\ref{TH02} we have,
\begin{eqnarray}
	H_{K} &=& \sum_{i=1}^{n} 
	   \left( 
	      \frac{1}{2} ||{\bf \tilde r_i}||^2 \left[\frac{m_0 + m_i}{m_0 m_i}\right] 
	      - G\frac{m_0 m_i}{||{\bf r_i}||} 
	   \right),\label{Ka}
	\\ 
	T_1 &=& \sum_{0 < i < j\leq n} \frac{\bf \tilde r_i \cdot \tilde r_j}{m_0},\label{T1a}
	\\
	U_1 &=& - G \sum_{0 < i < j\leq n} \frac{m_i m_j}{\Delta_{ij}}.\label{U1a}
\end{eqnarray}
The main advantage of this way to split the equations is that Kepler's third law is 
satisfied for the individual planets: $n^2a^3 = G(m_0 + m_i)$. But $H_I = T_1 + U_1$ is 
not integrable and $\{T_1, U_1\} \neq 0$. 

Using Eq.~\ref{TH01} we have the splitting introduced by \cite{Cham99},
\begin{eqnarray}
	H_{K}^* &=& \sum_{i=1}^{n} 
   \left( 
   \frac{1}{2} \frac{||{\bf \tilde r_i}||^2} {m_i} - G\frac{m_0 m_i}{||{\bf r_i}||} 
   \right),\label{Kb}
\\ 
	T_1^* &=& \frac{1}{2} \frac{|| \sum_{i=1}^n {\bf \tilde r_i} ||^2}{m_0}, \label{T1b}
\\
	U_1^* &=& - G \sum_{0 < i < j\leq n} \frac{m_i m_j}{\Delta_{ij}}.\label{U1b}
\end{eqnarray}
With this way to split the equations the mass parameter for the Keplerian orbits is 
$\mu = Gm_0$ for all of the planets. On the other hand, $T_1^*$ and $U_1^*$ 
(Eqs.~\ref{T1b}-\ref{U1b}) 
commute, i.e. $\{T^*_1, U^*_1\} = 0$ and this is a advantage when we build high
order splitting schemes. 
For simplicity let us consider $H_B = T_1$ and $H_C = U_1$ for both 
expressions. We recall that in Heliocentric coordinates we need to integrate
$\exp(\tau(B + C))$. Using Chambers' splitting (Eqs.~\ref{T1b}-\ref{U1b})
we have: 
\begin{equation}
\exp(\tau (B + C)) = \exp(\tau B)\exp(\tau C),
\end{equation}
which can be computed exactly and does not introduce any extra error terms to 
the splitting schemes discussed in Section~\ref{SymIntTST}. Instead using the first 
splitting expressions (Eqs.~\ref{T1a}-\ref{U1a}) we used  
\begin{equation}
\exp(\tau (B + C)) \approx
\exp(\frac{\tau}{2} C)\exp(\tau B)\exp(\frac{\tau}{2} C),
\end{equation}
and introduced error terms of order $\eps^3\tau^2$. To deal with this in 
Section~\ref{SymIntHelio} derived splitting schemes where an extra stage was added 
to get rid of these extra error terms. 

\subsection{Comparisons between the expressions} 

We recall that when we use Chambers expression (Eqs.~\ref{Kb}-\ref{U1b}) 
we use the splitting schemes discussed in Section~\ref{SymIntTST} with
\begin{equation}
\mathcal{S}(\tau) = 
   \prod_{i = 1}^n \exp(a_i\tau L_{H_{K^*}}) \exp(b_i\tau L_{T_1^*})\exp(b_i\tau L_{U_1^*}).  
\label{SymHelioDem}
\end{equation}
While when we use the classical expression (Eqs.~\ref{Ka}-\ref{U1a}) we 
use the splitting schemes discussed in Section~\ref{SymIntHelio} with
\begin{equation}
\mathcal{S}(\tau) = 
   \prod_{i = 1}^n \exp(a_i\tau L_{H_{Kep}})
		\exp(b_i\frac{\tau}{2} L_{T_1})\exp(b_i\tau L_{U_1})\exp(b_i\frac{\tau}{2} L_{T_1}).
\label{SymHelioClas}
\end{equation}

We compare the $\mathcal{ABA}$ schemes of orders $(8,2)$, $(8,4)$ and
$(10,6,4)$ for both splitting expressions. We recall that the schemes of order
$(8,4)$ and $(10,6,4)$ that use the classical splitting
(Eqs.~\ref{Ka}-\ref{U1a}) have one more stage than the schemes used with
Chambers splitting (Eqs.~\ref{Kb}-\ref{U1b}).  

\begin{figure}[bth]
\includegraphics[width = 0.32\textwidth]{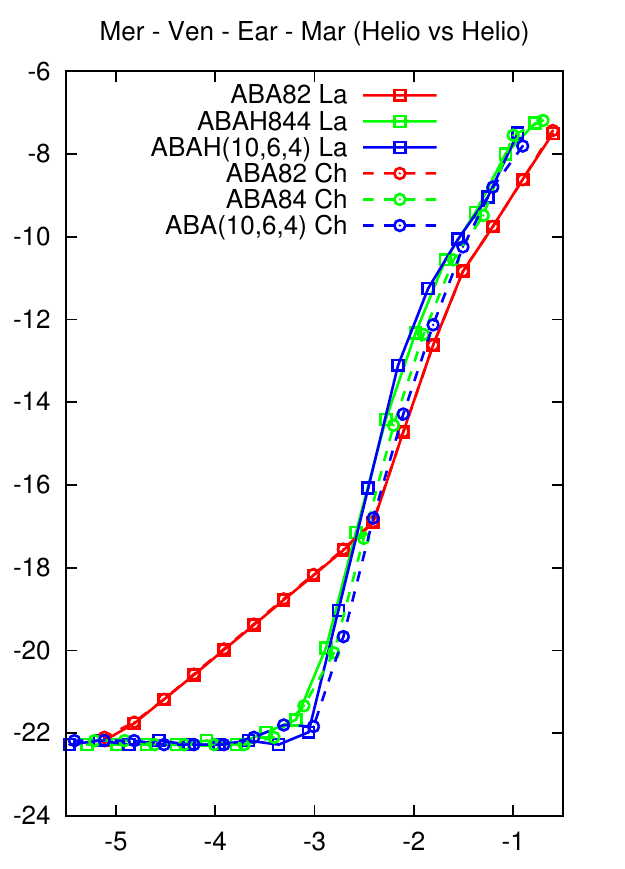}
\includegraphics[width = 0.32\textwidth]{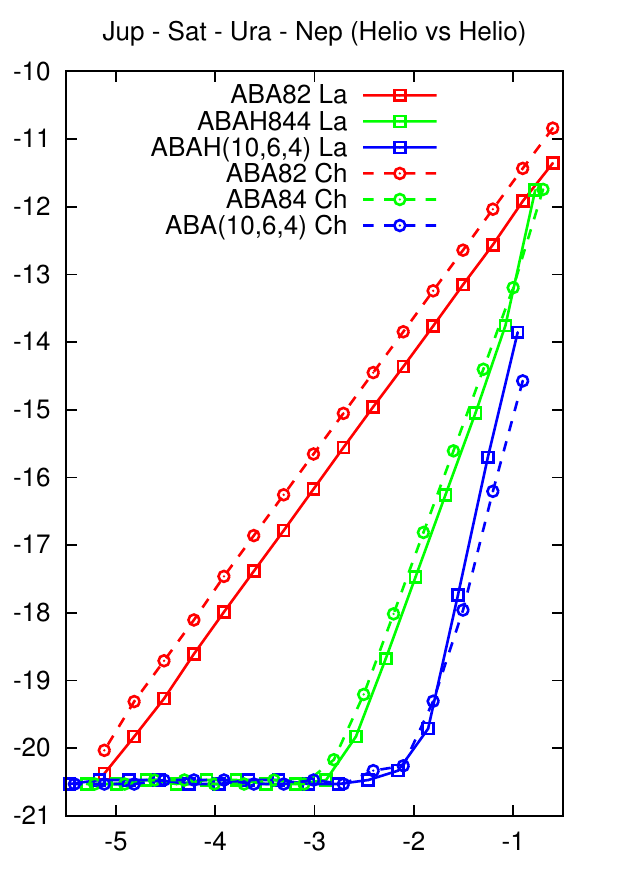}
\includegraphics[width = 0.32\textwidth]{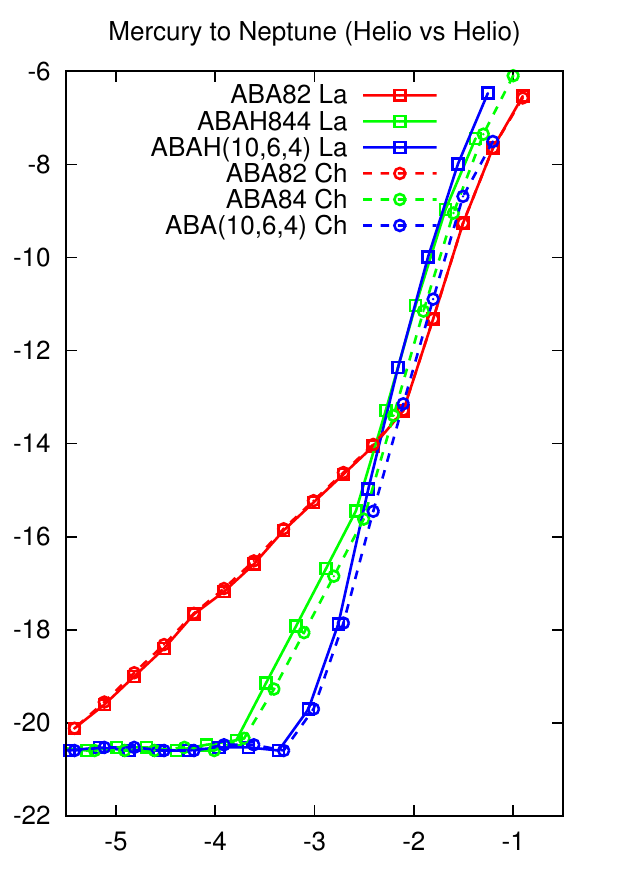}
\caption{ 
Comparison between the two expressions for Heliocentric coordinates: 
the classical expression (Eqs.~\ref{Ka}-\ref{U1a}) continuous 
lines and the Chambers expression (Eq.~\ref{Kb}-\ref{U1b}) discontinuous
lines. For the schemes $\mathcal{ABA}82$ (red), $\mathcal{ABA}84$ (green) and
$\mathcal{ABA}1064$ (blue). From left to right: the 4 inner planets, the 4 
outer planets and the whole Solar System. The $x$-axis represents the cost 
($\tau/s$) of the method and the $y$-axis the maximum energy variation for 
one integration with constant step-size $\tau$.
}
\label{FigHvsH}
\end{figure}

Figure~\ref{FigHvsH} summarise the performance of the different integrating 
schemes presented in Sections~\ref{SymIntTST} and~\ref{SymIntHelio}. From 
left to right we have the results for the inner planets, the outer planets
and the whole Solar System. The red lines show the performance of the 
$\mathcal{ABA}82$ scheme, the green lines are for the $\mathcal{ABA}84$ schemes 
and the blue lines are for the $\mathcal{ABA}1064$ schemes. We use 
continuous lines when we consider the classical splitting and discontinuous 
lines for Chambers splitting. 

As we can see, there is no significant difference between using one splitting
or the other. In some cases one is better that the other. The main advantage
that the splitting introduced by Chambers is that we do not require and extra
stage for a high-order scheme. 

\section{Comparison in Quadruple Precision} \label{Apndx_QP}

As we have discussed throughout the article, in many cases
we have seen that despite taking higher order methods no significant
improvement on the performance of the schemes was observed. This is the case of
the 4 inner planets in the Solar System, where the size of the perturbation is
so small that the extra stages to increase the order of the schemes are
useless. Here the round-off error dominates the terms in $\eps^2$ and $\eps^4$.
Similar results are also observed when we consider the whole Solar System. In order 
to see an improvement we need to use higher precision arithmetics. 
Here we have repeated the test from Sections~\ref{SymIntTST}
and~\ref{SymIntHelio} for the different integrating schemes using quadruple
precision arithmetics. We want to illustrate that the different schemes of
orders $(8,6,4)$ and $(10,6,4)$ perform better that those of order $(8,4)$.

\begin{figure}[!h]
\includegraphics[width = 0.32\textwidth]{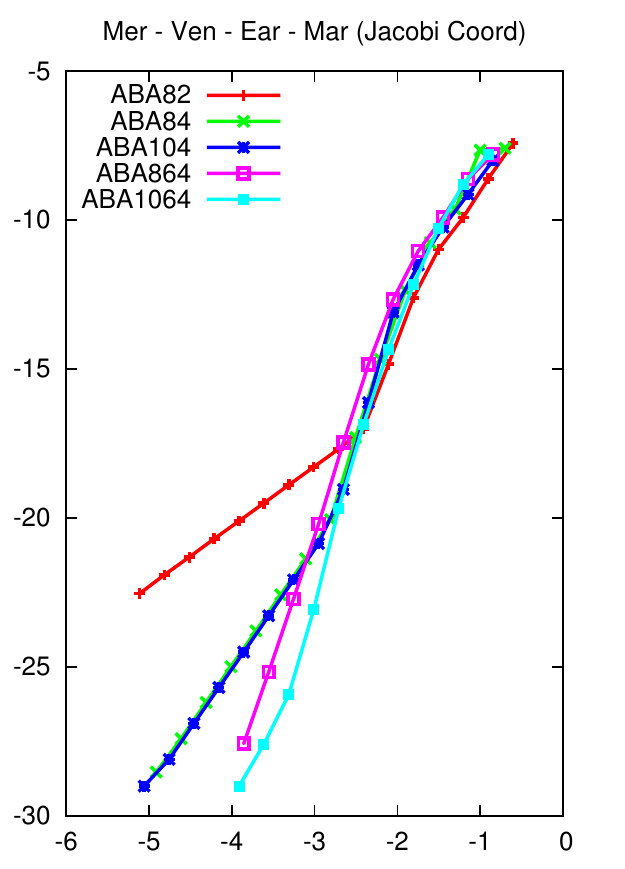}       
\includegraphics[width = 0.32\textwidth]{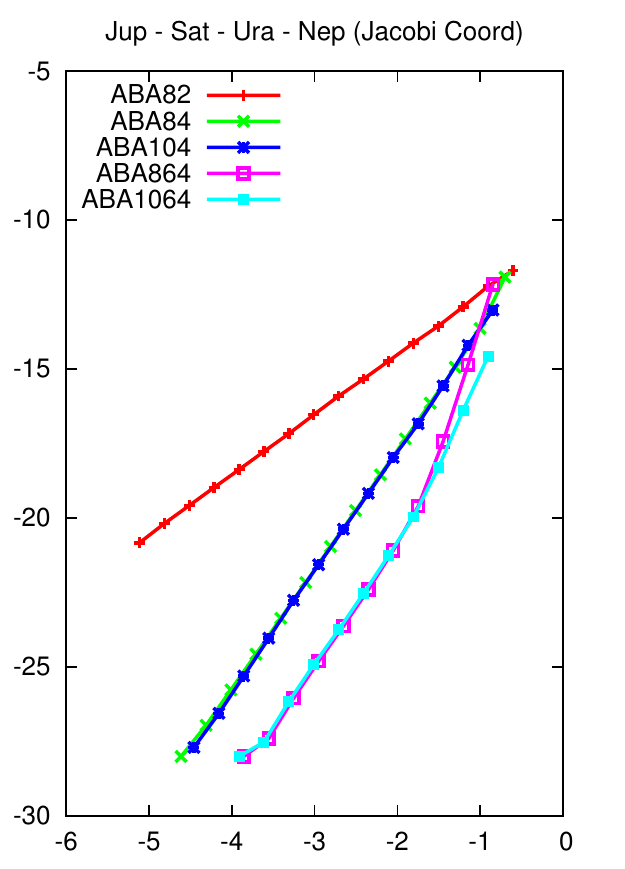} 
\includegraphics[width = 0.32\textwidth]{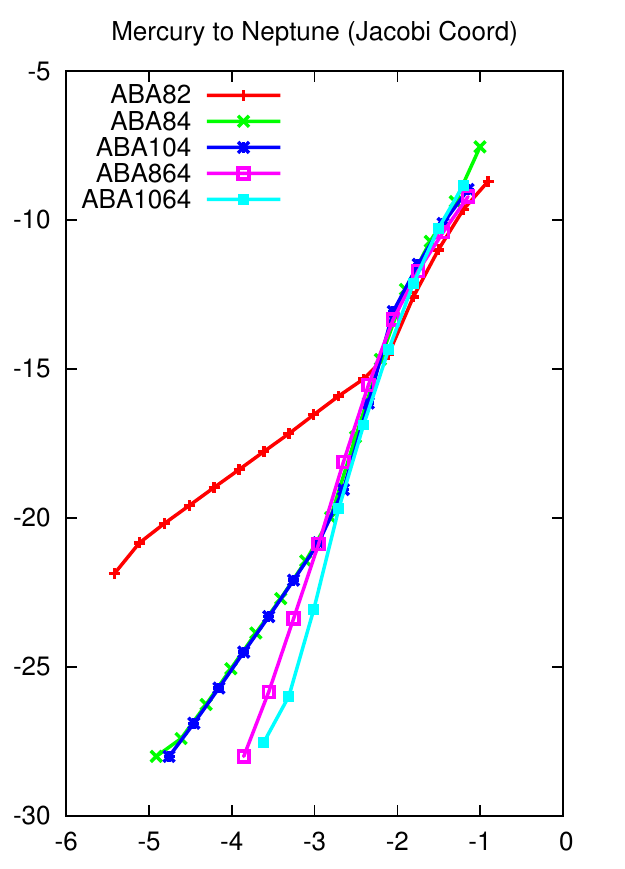}
\caption{
Comparison using Jacobi coordinates between the $\mathcal{ABA}82$,
$\mathcal{ABA}84$, $\mathcal{ABA}104$, $\mathcal{ABA}864$ and
$\mathcal{ABA}1064$ schemes using quadruple precision arithmetics. From left to right:
the 4 inner planets, the 4 outer planets and the whole Solar System. The
$x$-axis represents the cost $(\tau/s)$ and the $y$-axis the maximum energy
variation for one integration with constant step-size $\tau$.  
}\label{ABAJb_quad}
\end{figure}

\begin{figure}[!h]
\includegraphics[width = 0.32\textwidth]{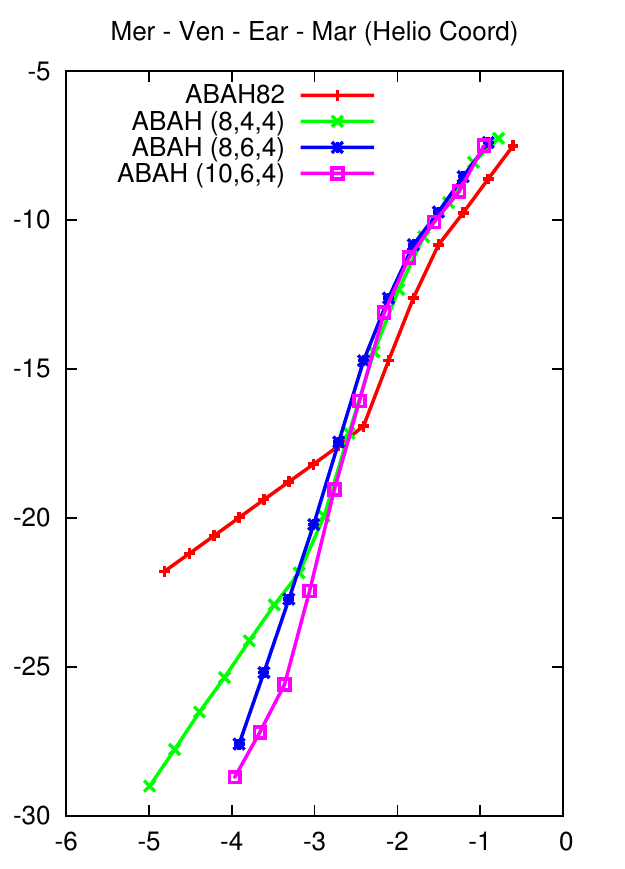}  
\includegraphics[width = 0.32\textwidth]{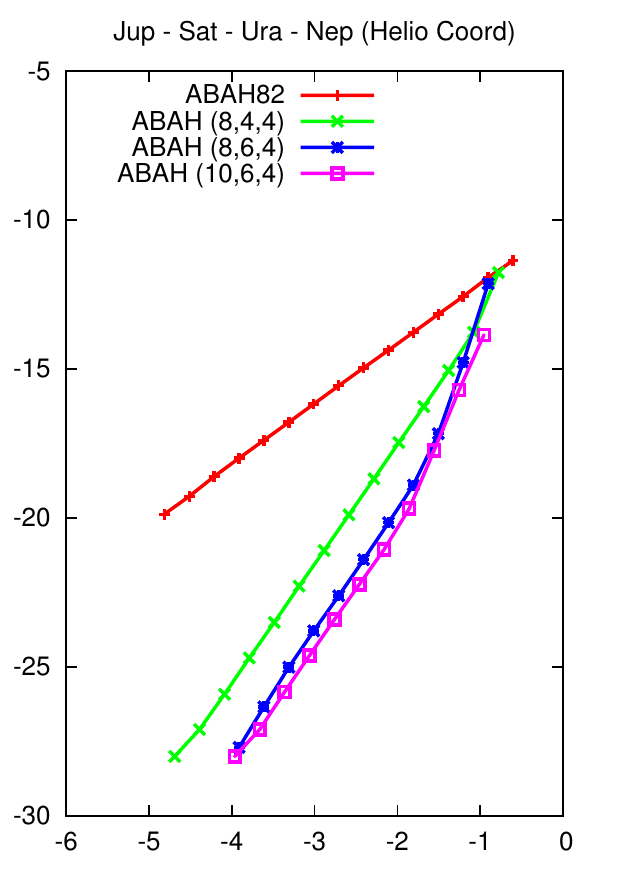}  
\includegraphics[width = 0.32\textwidth]{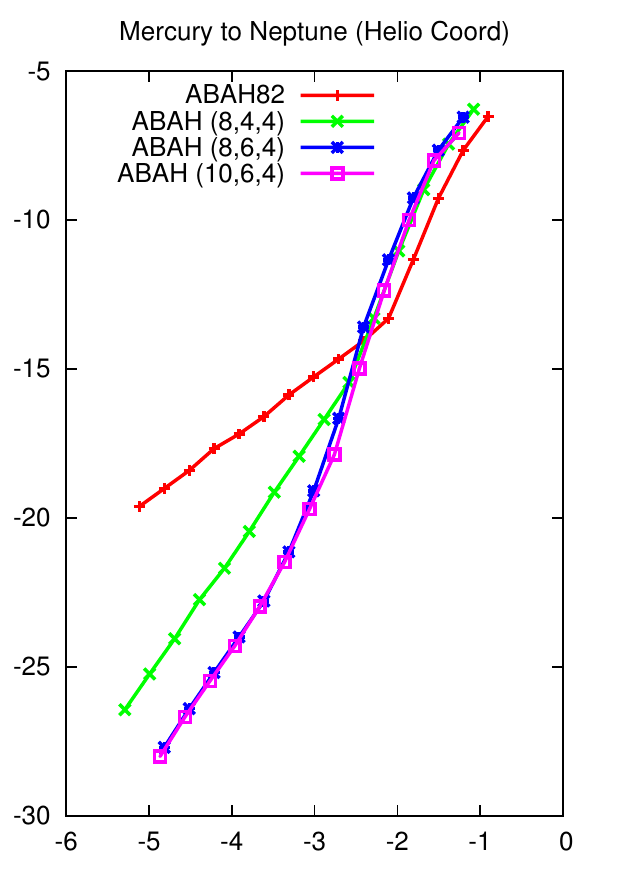}  
\caption{
Comparison using Heliocentric coordinates between the $\mathcal{ABAH}82$,
$\mathcal{ABAH}84$, $\mathcal{ABAH}864$ and $\mathcal{ABAH}1064$ schemes
using quadruple precision arithmetics. From left to right: the 4 inner
planets, the 4 outer planets and the whole Solar System. The $x$-axis
represents the cost $(\tau/s)$ and the $y$-axis the maximum energy variation
for one integration with constant step-size $\tau$.  
}\label{ABAHe_quad}
\end{figure}

In Figures~\ref{ABAJb_quad} and~\ref{ABAHe_quad} we show the results for the
same test models used throughout the article for Jacobi and Heliocentric
coordinates respectively using quadruple precision arithmetics. For Jacobi 
coordinates (Figure~\ref{ABAJb_quad}) we compare the $\mathcal{ABA}82$,
$\mathcal{ABA}84$, $\mathcal{ABA}104$, $\mathcal{ABA}864$ and
$\mathcal{ABA}1064$ schemes. For Heliocentric coordinates (Figure~\ref{ABAHe_quad}) 
we compare the $\mathcal{ABAH}82$, $\mathcal{ABAH}84$, $\mathcal{ABAH}864$ and 
$\mathcal{ABAH}1064$. 
As we can see in Figure~\ref{ABAJb_quad} for Jacobi coordinates, the
$\mathcal{ABA}864$ and $\mathcal{ABA}1064$ \citep{Bla12} do improve the
performance of the $\mathcal{ABA}84$ \citep{McLa9502}. Notice also that for the
4 inner planets (Figure~\ref{ABAJb_quad} left) and the whole Solar System
(Figure~\ref{ABAJb_quad} right) the improvement is achieved for small
step-sizes, where the energy variation is bellow the machines epsilon for extended 
arithmetics precision. In Figure~\ref{ABAHe_quad} similar results are observed
for Heliocentric coordinates. 

From these experiments we see how the $\mathcal{ABA}$ splitting methods of orders
$(8,6,4)$ and $(10,6,4)$ for both set of coordinates improve the performance of the 
\cite{McLa9502} $\mathcal{ABA}84$ and the \cite{LaRo01} $\mathcal{ABA}82$.

\end{appendix}

\clearpage
\begin{acknowledgements}
This work was supported by GTSNext project.
The work of SB, FC, JM and AM has been partially supported by Ministerio de
Ciencia e Innovaci\'on (Spain) under project MTM2010-18246-C03
(co-financed by FEDER Funds of the European Union).
\end{acknowledgements}

\bibliographystyle{spbasic}      
\bibliography{bib_nbp}   

\begin{thebibliography}{37}
\providecommand{\natexlab}[1]{#1}
\providecommand{\url}[1]{{#1}}
\providecommand{\urlprefix}{URL }
\expandafter\ifx\csname urlstyle\endcsname\relax
  \providecommand{\doi}[1]{DOI~\discretionary{}{}{}#1}\else
  \providecommand{\doi}{DOI~\discretionary{}{}{}\begingroup
  \urlstyle{rm}\Url}\fi
\providecommand{\eprint}[2][]{\url{#2}}

\bibitem[{Blanes et~al(2012)Blanes, Casas, Farr\'es, Laskar, Makazaga, and
  Murua}]{Bla12}
Blanes S, Casas F, Farr\'es A, Laskar J, Makazaga J, Murua A (2012) New
  families of symplectic splitting methods for numerical integration in
  dynamical astronomy. Submitted

\bibitem[{{Chambers}(1999)}]{Cham99}
{Chambers} JE (1999) {A hybrid symplectic integrator that permits close
  encounters between massive bodies}. Monthly Notices of the Royal Astronomical
  Society 304:793--799, \doi{10.1046/j.1365-8711.1999.02379.x}

\bibitem[{Chambers and Murison(2000)}]{Cham00}
Chambers JE, Murison MA (2000) Pseudo-high-order symplectic integrators. The
  Astronomical Journal 119(1):425

\bibitem[{Danby(1992)}]{Danby92}
Danby JMA (1992) Fundamentals of Celestial Mechanics, Willmann-Bell, p 483

\bibitem[{{Duncan} et~al(1998){Duncan}, {Levison}, and {Lee}}]{DuLeLee98}
{Duncan} MJ, {Levison} HF, {Lee} MH (1998) {A Multiple Time Step Symplectic
  Algorithm for Integrating Close Encounters}. Astronomical Journal
  116:2067--2077, \doi{10.1086/300541}

\bibitem[{Goldman and Kaper(1996)}]{goldman96noo}
Goldman D, Kaper T (1996) ${N}$th-order operator splitting schemes and
  nonreversible systems. SIAM J Numer Anal 33:349--367

\bibitem[{Hairer et~al(2006)Hairer, Lubich, and Wanner}]{hairer06gni}
Hairer E, Lubich C, Wanner G (2006) Geometric Numerical Integration.
  Structure-Preserving Algorithms for Ordinary Differential Equations, {S}econd
  edn. Springer-Verlag

\bibitem[{Kahan(1965)}]{Kahan}
Kahan W (1965) Pracniques: further remarks on reducing truncation errors.
  Commun ACM 8:40--, \doi{10.1145/363707.363723}

\bibitem[{{Kinoshita} et~al(1991){Kinoshita}, {Yoshida}, and
  {Nakai}}]{KiYoNa91}
{Kinoshita} H, {Yoshida} H, {Nakai} H (1991) {Symplectic integrators and their
  application to dynamical astronomy}. Celestial Mechanics and Dynamical
  Astronomy 50:59--71

\bibitem[{Koseleff(1993)}]{Kose93}
Koseleff PV (1993) Relations among lie formal series and construction of
  symplectic integrators. In: Cohen~G MT, O M (eds) in Applied Algebra,
  Algebraic Algorithms and Error Correcting Codes (AAECC-10), Springer Verlag,
  New York, Lect. Not. Comp. Sci, vol 673, pp 213--230

\bibitem[{Koseleff(1996)}]{Kose96}
Koseleff PV (1996) Exhaustive search of symplectic integrators using computer
  algebra. Fields Institute Communications 10

\bibitem[{Laskar(1989)}]{Lask1989a}
Laskar J (1989) A numerical experiment on the chaotic behaviour of the solar
  system. Nature 338:237,
  \urlprefix\url{http://adsabs.harvard.edu/abs/1989Natur.338..237L}

\bibitem[{Laskar(1990{\natexlab{a}})}]{Lask1990a}
Laskar J (1990{\natexlab{a}}) The chaotic motion of the solar system - a
  numerical estimate of the size of the chaotic zones. Icarus 88:266--291,
  \urlprefix\url{http://adsabs.harvard.edu/abs/1990Icar...88..266L}

\bibitem[{Laskar(1990{\natexlab{b}})}]{CursLaskar}
Laskar J (1990{\natexlab{b}}) Les M\'ethodes Modernes de la Mec\'anique
  C\'eleste (Goutelas, France, 1989), Editions Fronti\`eres, chap Syst\`emes de
  Variables et El\'ements, pp 63--87

\bibitem[{Laskar and Robutel(2001)}]{LaRo01}
Laskar J, Robutel P (2001) High order symplectic integrators for perturbed
  hamiltonian systems. Celestial Mechanics and Dynamical Astronomy 80:39--62,
  10.1023/A:1012098603882

\bibitem[{Laskar et~al(1992)Laskar, Quinn, and Tremaine}]{LaskQuin1992a}
Laskar J, Quinn T, Tremaine S (1992) Confirmation of resonant structure in the
  solar system. Icarus 95:148--152, \doi{DOI: 10.1016/0019-1035(92)90196-E},
  \urlprefix\url{http://adsabs.harvard.edu/abs/1992Icar...95..148L}

\bibitem[{Laskar et~al(2004)Laskar, Robutel, Joutel, Gastineau, Correia, and
  Levrard}]{LaskRobu2004a}
Laskar J, Robutel P, Joutel F, Gastineau M, Correia ACM, Levrard B (2004) A
  long-term numerical solution for the insolation quantities of the earth.
  Astronomy and Astrophysics 428:261--285

\bibitem[{Laskar et~al(2011{\natexlab{a}})Laskar, Fienga, Gastineau, and
  Manche}]{LaskFien2011a}
Laskar J, Fienga A, Gastineau M, Manche H (2011{\natexlab{a}}) La2010: a new
  orbital solution for the long-term motion of the earth. Astronomy and
  Astrophysics 532:89, \doi{DOI: 10.1051/0004-6361/201116836; eprintid:
  arXiv:1103.1084}

\bibitem[{Laskar et~al(2011{\natexlab{b}})Laskar, Gastineau, Delisle,
  Farr{\'e}s, and Fienga}]{LaskGast2011a}
Laskar J, Gastineau M, Delisle JB, Farr{\'e}s A, Fienga A (2011{\natexlab{b}})
  Strong chaos induced by close encounters with ceres and vesta. Astronomy and
  Astrophysics 532:L4, \doi{DOI: 10.1051/0004-6361/201117504}

\bibitem[{Lourens et~al(2004)Lourens, Hilgen, Laskar, Shackleton, and
  Wilson}]{L.J.Wils2004a}
Lourens L, Hilgen F, Laskar J, Shackleton N, Wilson D (2004) The neogene
  period. In: Gradstein F, Ogg J, Smith A (eds) A Geological Timescale 2004, pp
  409--440

\bibitem[{McLachlan and Quispel(2002)}]{mclachlan02sm}
McLachlan R, Quispel R (2002) Splitting methods. Acta Numerica 11:341--434

\bibitem[{McLachlan(1995)}]{McLa9502}
McLachlan RI (1995) Composition methods in the presence of small parameters.
  BIT Numerical Mathematics 35:258--268, 10.1007/BF01737165

\bibitem[{McLachlan(2002)}]{McLa02}
McLachlan RI (2002) Families of high-order composition methods. Numerical
  Algorithms 31:233--246

\bibitem[{Milankovitch(1941)}]{Mila1941a}
Milankovitch M (1941) Kanon der Erdbestrahlung und seine Anwendung auf das
  Eiszeitenproblem. Spec. Acad. R. Serbe, Belgrade

\bibitem[{Morbidelli(2002)}]{Morb2002a}
Morbidelli A (2002) Modern integrations of solar system dynamics. Annual Review
  of Earth and Planetary Sciences 30:89--112, \doi{DOI:
  10.1146/annurev.earth.30.091201.140243},
  \urlprefix\url{http://adsabs.harvard.edu/abs/2002AREPS..30...89M}

\bibitem[{Murua and Sanz-Serna(1999)}]{Murua15041999}
Murua A, Sanz-Serna J (1999) Order conditions for numerical integrators
  obtained by composing simpler integrators. Philosophical Transactions of the
  Royal Society of London Series A: Mathematical, Physical and Engineering
  Sciences 357(1754):1079--1100, \doi{10.1098/rsta.1999.0365},
  \urlprefix\url{http://rsta.royalsocietypublishing.org/content/357/1754/1079.%
abstract},
  \eprint{http://rsta.royalsocietypublishing.org/content/357/1754/1079.full.pd%
f+html}

\bibitem[{Quinn et~al(1991)Quinn, Tremaine, and Duncan}]{QuinTrem1991a}
Quinn TR, Tremaine S, Duncan M (1991) A three million year integration of the
  earth's orbit. The Astronomical Journal 101:2287--2305,
  \urlprefix\url{http://adsabs.harvard.edu/abs/1991AJ....101.2287Q}

\bibitem[{{Saha} and {Tremaine}(1994)}]{SaTre94}
{Saha} P, {Tremaine} S (1994) {Long-term planetary integration with individual
  time steps}. Astronomical Journal 108:1962--1969, \doi{10.1086/117210},
  \eprint{arXiv:astro-ph/9403057}

\bibitem[{Sheng(1989)}]{sheng89slp}
Sheng Q (1989) Solving linear partial differential equations by exponential
  splitting. IMA J Numer Anal 9:199--212

\bibitem[{Sussman and Wisdom(1992)}]{SussWisd1992a}
Sussman GJ, Wisdom J (1992) Chaotic evolution of the solar system. Science
  257:56--62, \urlprefix\url{http://adsabs.harvard.edu/abs/1992Sci...257...56S}

\bibitem[{Suzuki(1990)}]{Suzu90}
Suzuki M (1990) Fractal decomposition of exponential operators with
  applications to many-body theories and monte carlo simulations. Physics
  Letters A 146(6):319 -- 323, \doi{10.1016/0375-9601(90)90962-N}

\bibitem[{Suzuki(1991)}]{Suzu91}
Suzuki M (1991) General theory of fractal path integrals with applications to
  many-body theories and statistical physics. Journal of Mathematical Physics
  32(2):400--407

\bibitem[{Viswanath(2002)}]{Visw02}
Viswanath D (2002) {How Many Timesteps for a Cycle? Analysis of the
  Wisdom-Holman Algorithm}. BIT Numerical Mathematics 42:194--205

\bibitem[{Wisdom(2006)}]{Wisd06}
Wisdom J (2006) Symplectic correctors for canonical heliocentric n-body maps.
  The Astronomical Journal 131(4):2294

\bibitem[{{Wisdom} and {Holman}(1991)}]{WiHo91}
{Wisdom} J, {Holman} M (1991) {Symplectic maps for the n-body problem}.
  Astronomical Journal 102:1528--1538, \doi{10.1086/115978}

\bibitem[{{Wisdom} et~al(1996){Wisdom}, {Holman}, and {Touma}}]{WiHoTo96}
{Wisdom} J, {Holman} M, {Touma} J (1996) {Symplectic Correctors}. Fields
  Institute Communications, Vol~10, p~217 10:217--+

\bibitem[{{Y}oshida(1990)}]{Yosh90}
{Y}oshida H (1990) Construction of higher order symplectic integrators. Physics
  Letters A 150(5-7):262 -- 268, \doi{10.1016/0375-9601(90)90092-3}

\end{thebibliography}

\end{document}